\documentclass[showpacs,aps,epsfig,nofootinbib]{revtex4}
\usepackage{mathrsfs}

\usepackage{graphicx}
\newcommand{\ba}{\begin{eqnarray}} \newcommand{\ea}{\end{eqnarray}}
\usepackage{amsfonts}
\usepackage{epstopdf}
\usepackage{latexsym}
\usepackage{amssymb}
\usepackage{amssymb}
\usepackage{color}
\usepackage{ulem}

\usepackage[center]{subfigure}

\begin{document}

 \newcommand{\bq}{\begin{equation}}
 \newcommand{\eq}{\end{equation}}
 \newcommand{\bqn}{\begin{eqnarray}}
 \newcommand{\eqn}{\end{eqnarray}}
 \newcommand{\nb}{\nonumber}
 \newcommand{\lb}{\label}
\newcommand{\PRL}{Phys. Rev. Lett.}
\newcommand{\PL}{Phys. Lett.}
\newcommand{\PR}{Phys. Rev.}
\newcommand{\CQG}{Class. Quantum Grav.}

\title{Gravitational Quasinormal Modes of Regular Phantom Black Hole}

\author{Jin Li}
\email{cqstarv@hotmail.com}
\affiliation {College of Physics, Chongqing University,
Chongqing 401331, China}
\affiliation {State Key Laboratory Theoretical Physics, Institute of Theoretical Physics, Chinese Academy of Sciences, Beijing 100190, China}
\
\author{Kai Lin}
\email{lk314159@hotmail.com}
\affiliation {Universidade Federal de Itajub\'a, Instituto de F\'isica e Qu\'imica, CEP 37500-903, Itajub\'a, Brazil}
\affiliation {Department of Astronomy, China West Normal University, Nanchong, Sichuan 637002, China}
\
\author{Hao Wen}
\email{wenhao@cqu.edu.cn}
\affiliation {College of Physics, Chongqing University,
Chongqing 401331, China}
\
\author{Wei-Liang Qian}
\email{wlqian@usp.br}
\affiliation{Escola de Engenharia de Lorena, Universidade de S\~ao Paulo, S\~ao Paulo, SP, Brasil}
\affiliation{Faculdade de Engenharia de Guaratinguet\'a, Universidade Estadual Paulista, Guaratinguet\'a, SP, Brasil}
\date{\today}

\begin{abstract}
We investigate the gravitational quasi-normal modes (QNMs) for a type of regular black hole (BH) known as phantom BH, which is a static self-gravitating solution of a minimally coupled phantom scalar field with a potential. 
The studies are carried out for three different spacetimes: asymptotically flat, de Sitter (dS), and anti de Sitter (AdS). 
In order to consider the standard odd parity and even parity of gravitational perturbations, the corresponding master equations are derived.
The QNMs are discussed by evaluating the temporal evolution of the perturbation field which, in turn, provides direct information on the stability of BH spacetime. 
It is found that in asymptotically flat, dS and AdS spacetimes, the gravitational perturbations have similar characteristics for both odd and even parities. 
The decay rate of perturbation is strongly dependent on the scale parameter $b$, which measures the coupling strength between phantom scalar field and the gravity. 
Furthermore, through the analysis of Hawking radiation, it is shown that the thermodynamics of such regular phantom BH is also influenced by $b$. 
The obtained results might shed some light on the quantum interpretation of QNM perturbation.

\end{abstract}


\maketitle
\section{introduction}

As a major topic in cosmology, the accelerated expansion of our universe has caused widespread concern in the scientific community. 
Since the effect of gravity causes the expansion speed to slow down, the accelerated expansion of the universe implies the existence of an unknown form of energy in the universe. 
The latter provides a repulsive force to push the expansion of the universe. Such unknown energy is called dark energy (DE). 
Subsequently, a large number of DE models have been proposed, among which the one with cosmological constant is the most famous. 
Even though the model of DE with the cosmological constant is reasonable in physical theory and consistent with most observations, 
two difficulties still remain unsolved, namely, how to derive ``vacuum energy" from quantum field theory and why the magnitude of present DE and dark matter are of the same order.

Many modern astrophysics observations indicated the possibility of pressure to density ratio $w<-1$. 
For example, a model-free data analysis from 172 type Ia supernovae (SNIa) resulted in a range of $-1.2<w<-1$ for our present epoch \cite{prl5}. 
According to the WMAP data during 7 years, $w=-1.10^{+0.14}_{-0.14}(1\sigma)$ \cite{ins2}. 
By using the data from Chandra telescope, an analysis of the hot gas in 26 X-ray luminous dynamically relaxed galaxy clusters gives $w=-1.20^{+0.24}_{-0.28}$ \cite{prl6}. 
The data on SNIa from the SNLS3 sample estimates $w=-1.069^{+0.091}_{-0.092}$ \cite{ins3}. 
In fact, several DE models with a super-negative equation of state provide better fits to the above data \cite{r1add1,r1add2,r1add3,r1add5}. 
And all these approaches are in favor of phantom DE scenario \cite{DEM1,DEM2,DEM3,DEM4,DEM5}, in which a constant equation of state parameter is used \cite{r1add6,r1add7}. 
This implies that the phantom model might be meaningful for in-depth understanding of DE. 

In the phantom model, the signature of the metric is $+2$, and the action of the model reads
\begin{equation}
S=\int\sqrt{-g}d^{4}x[R+\epsilon g^{\mu\nu}\partial_{\mu}\phi\partial_{\nu}\phi-2V(\phi)],
\label{action}
\end{equation}
where $R$ is the scalar curvature, $V(\phi)$ is the potential of the scalar field, $\epsilon=-1$ corresponds to a phantom scalar field while $\epsilon=+1$ is for a normal canonical scalar field. 

Bronnikov and Fabris first investigated the properties of BH with phantom scalar field in vacuum, and derived a phantom regular BH solution 10 years ago\cite{mainref2}. 
Inside the event horizon of such phantom BH there is no singularity similar to the case of regular BHs with nonlinear electrodynamics sources\cite{regularQNMs}.
Outside the event horizon, the properties of a phantom BH are similar to those of a Schwarzschild BH. 
Due to the absence of the singularity, such phantom regular BH solution has attracted much attention from researchers.

On the other hand, the research of BH perturbation has always been an important issue in BH physics. 
The first work on QNM in AdS spacetime was about scalar wave in Schwarzschild-AdS spacetime\cite{AdS0}, which is then followed by a study on scalar wave in topological AdS spacetime \cite{AdS00}. 
There are a large number of works on regular BH's QNMs\cite{mainref,regularQNMs,regularQNMs2,regularQNMs3,regularQNMs4,finit4,Hawking8}. 
Among various types of perturbation, gravitational perturbation is generally considered to be the most important form due to its practical significance. 
The intrinsic properties and the stability of a BH can be unfolded through its corresponding gravitational perturbation. 
In the fifties of last century, Regge and Wheeler began to study the gravitational perturbations of static spherically symmetric BHs.
It was pointed out later that \cite{HD1} the higher dimensional gravitational perturbations can be classified into three types, namely, scalar-gravitational, vector-gravitational, and tensor-gravitational perturbations.
The first two types are associated to odd (vector-gravitational) and even (scalar-gravitational) parity in accordance with the spatial inversion symmetry of the perturbations, and are of great physical interest \cite{wheeler}.
These findings significantly simplify the study of gravitational perturbation of BH. 
Subsequently, people developed many new methods, and further studies on the gravitational perturbations result in a large number of master equations for various forms of BHs 
in 4-dimensions \cite{wheeler,Gq4D1,Gq4D2,Gq4D3}, in higher dimensions \cite{HD1}, and for stationary BHs \cite{Gq1,Gq2}.
In fact, gravitational perturbations of a BH may generate relatively strong gravitational waves (GWs). 
Recently, the GWs from a binary BH system has been detected by LIGO \cite{LIGO}, so BH is proven to be the most probable source of GWs by modern technology. 
Meanwhile since many alternative theories of gravity can produce the same GW signal within the present accuracy in far field, the reported GW detection still leaves a window for alternative gravity theories \cite{PLBnew}, which includes the theory of phantom BHs. 
Therefore, the properties of QNMs of gravitational perturbation near the horizon of phantom BH may provide us essential information on the underlying physics of gravity theory. 
This is the main purpose of the present study.

The thermodynamics of BHs is also an important subject in BH physics. 
Some works indicated that Hawking radiation can be considered as an effective quantum thermal radiation around the horizon \cite{Hawking1,Hawking2}, 
where the corresponding Hawking temperature can be derived from the tunneling rate \cite{Hawking1,Hawking2,Hawking3,Hawking4,Hawking5,Hawking6}. 
Furthermore, a natural correspondence between Hawking radiation and QNM has been established recently \cite{Hawking1,Hawking2,Hawking3,Hawking5,Hawking7}. 
Therefore, in this work, we will also investigate the Hawking radiation of regular phantom BH.

The paper is organized as follows. 
In section \ref{2}, we briefly review the regular phantom BH solutions and discuss their properties in three different spacetimes, namely, asymptotically flat, de Sitter (dS) and anti de Sitter (AdS).
In this work, we focus on the odd parity and even parity gravitational perturbations. 
As the main component of this paper, section \ref{3} includes two subsections. 
In subsection A, we derive the master equation for odd parity gravitational perturbation, and analyze the corresponding temporal evolution of the perturbed metric; 
In subsection B, corresponding studies are carried out for the even parity gravitational perturbation. 
In section \ref{4}, we calculate the Hawking radiation of the regular phantom BH. 
We summarize our results and draw concluding remarks in section \ref{5}.

\section{The general metric for regular phantom black holes}
\label{2}
In this section, we discuss the phantom ($\epsilon=-1$) regular BH solution by considering the following static metric with spherical symmetry
\begin{equation}
ds^{2}=-f(r)dt^{2}+\frac{dr^{2}}{f(r)}+p(r)^{2}(d\theta^{2}+\rm{sin}^{2}\theta d\varphi^{2}).
\label{metric}
\end{equation}
According to the action, Eq.(\ref{action}), the field equation for a self-gravitating minimally coupled scalar field with an arbitrary potential $V (\phi)$ can be expressed as
\begin{equation}
\widehat{G}_{\mu\nu}=R_{\mu\nu}-\frac{g_{\mu\nu}}{2}(R-\epsilon \phi^{;\alpha}\phi_{;\alpha}-2V(\phi))-\epsilon\phi_{;\mu}\phi_{;\nu}=0.
\label{fieldequation}
\end{equation}
By combining the scalar field equation,
\begin{equation}
\epsilon \phi^{;\alpha}_{~;\alpha}-\frac{dV(\phi)}{d\phi}=0,
\end{equation}
a regular phantom BH solution can be obtained as
\begin{equation}
f(r)=\left\{\frac{c}{b^{2}}+\frac{1}{p^{2}(r)}+\frac{3m}{b^{3}}\left[\frac{br}{p^{2}(r)}+\text{Arctan}(\frac{r}{b})\right]\right\}p^{2}(r),
\label{fr}
\end{equation}
where
\begin{equation}
p(r)=\sqrt{b^{2}+r^{2}},
\label{pr}
\end{equation}
and
\begin{equation}
V(\phi(r))=-\frac{c}{b^{2}}\frac{p^{2}+2r^{2}}{p^{2}}-\frac{3m}{b^{3}}\left\{\frac{3br}{p^{2}}+\frac{p^{2}+2r^{2}}{p^{2}}\text{Arctan}(\frac{r}{b})\right\},~~~~~\phi(r)=\sqrt{2}\epsilon \text{Arctan}(\frac{r}{b})+\phi_{0},
\label{Vr}
\end{equation}
$m$ is the Schwarzschild mass defined in the usual way, $c$ and $b$ are integration constant and scale parameter respectively. Then it is necessary to determine the possible kinds of spacetime for such phantom BH, which can be classified as a regular infinity $(r\rightarrow\infty)$ to be flat, de Sitter (dS) or Anti-de Sitter (AdS). The corresponding parameters $c,b,m$ should be restricted in each spacetime.

\textit{For the asymptotically flat spacetime}, in accordance with Eq.(\ref{fr}), one has $c=-3\pi m/2b$ and
\begin{equation}
m=\frac{2b^{3}}{3\left[\pi b^{2}-2br_{\text{h}}+\pi r_{\text{h}}^{2}-2b^{2}\text{Arctan}(r_{\text{h}}/b)-2r_{\text{h}}^{2}\text{Arctan}(r_{\text{h}}/b)\right]}.
\end{equation}
In this case, the spacetime is asymptotically flat, namely, $r\rightarrow\infty$,$f(r)\rightarrow1$. And $r_{\text{h}}$ is the event horizon of the phantom BH. We note when $b\rightarrow 0$, Eq.(\ref{fr}) becomes Schwarzschild flat spacetime. 

\textit{For the de Sitter spacetime}, one has
\begin{equation}
c=-\frac{\text{b}^2 \left[\left(\text{b}^2+r_{\text{c}}^2\right) \text{Arctan}\left(\frac{r_{\text{c}}}{\text{b}}\right)-\left(\text{b}^2+r_{\text{h}}^2\right) \text{Arctan}\left(\frac{r_{\text{h}}}{\text{b}}\right)+\text{b} (r_{\text{c}}-r_{\text{h}})\right]}{\left(\text{b}^2+r_{\text{c}}^2\right) \left(\text{b}^2+r_{\text{h}}^2\right) \text{Arctan}\left(\frac{r_{\text{c}}}{\text{b}}\right)-\left(\text{b}^2+r_{\text{c}}^2\right) \left(\text{b}^2+r_{\text{h}}^2\right) \text{Arctan}\left(\frac{r_{\text{h}}}{\text{b}}\right)+\text{b} (r_{\text{c}}-r_{\text{h}}) \left(\text{b}^2-r_{\text{c}} r_{\text{h}}\right)},
\end{equation}
\begin{equation}
m=\frac{\text{b}^3 \left(r_{\text{c}}^2-r_{\text{h}}^2\right)}{3\left[\left(\text{b}^2+r_{\text{c}}^2\right) \left(\text{b}^2+r_{\text{h}}^2\right) \text{Arctan}\left(\frac{r_{\text{c}}}{\text{b}}\right)-\left(\text{b}^2+r_{\text{c}}^2\right) \left(\text{b}^2+r_{\text{h}}^2\right) \text{Arctan}\left(\frac{r_{\text{h}}}{\text{b}}\right)+\text{b} (r_{\text{c}}-r_{\text{h}}) \left(\text{b}^2-r_{\text{c}} r_{\text{h}}\right)\right]},
\end{equation}
where $r_{\text{c}}$,  $r_{\text{h}}$ is the cosmological horizon and  event horizon respectively. We note when $b\rightarrow 0$, Eq.(\ref{fr}) becomes Schwarzschild dS metric.

\textit{For the Anti-de sitter spacetime}, we choose $f(r)\rightarrow r^{2}$ with $r\rightarrow \infty$ without loss of generality. By expanding the Eq.(\ref{fr}) around infinity, one finds
\begin{equation}
c=\frac{2b^{3}-3\pi m}{2b},
\label{adsc}
\end{equation}
and
\begin{equation}
m=\frac{2 \text{b}^3 \left(\text{b}^2+r_{\text{h}}^2+1\right)}{3\left[\pi  \text{b}^2-2 \text{b}^2 \text{Arctan}\left(\frac{r_{\text{h}}}{\text{b}}\right)-2 r_{\text{h}}^2 \text{Arctan}\left(\frac{r_{\text{h}}}{\text{b}}\right)-2 \text{b}r_{\text{h}}+\pi  r_{\text{h}}^2\right]},
\label{adsr0}
\end{equation}
where $r_{\text{h}}$ is the event horizon of AdS spacetime. 
We note when $b\rightarrow 0$, Eq.(\ref{fr}) can be returned to Schwarzschild AdS spacetime. 
In this context, the parameter $b$ measures the coupling strength between phantom scalar field and the gravity for all three spacetimes. 

\begin{figure}
\includegraphics[height=4cm]{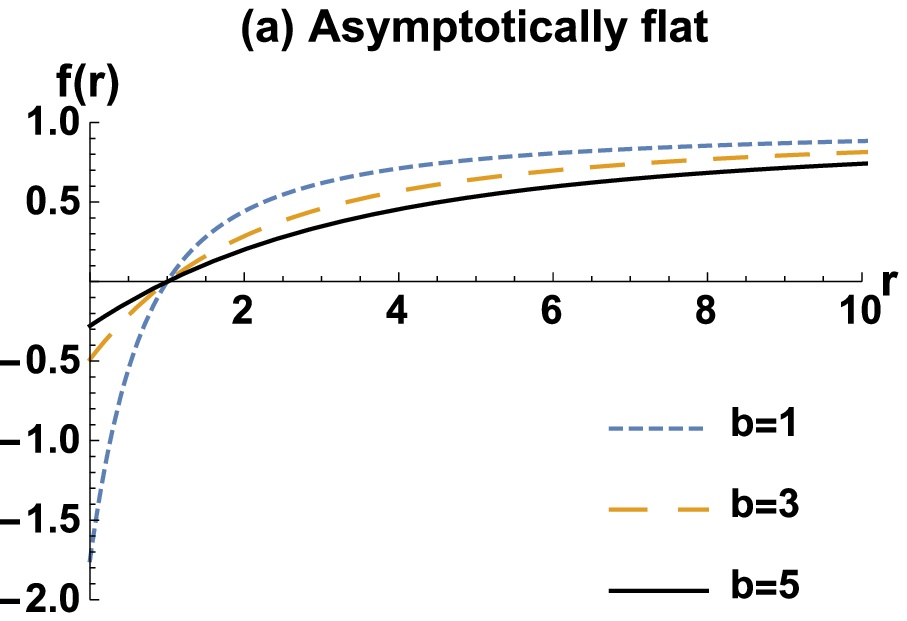}\includegraphics[height=4cm]{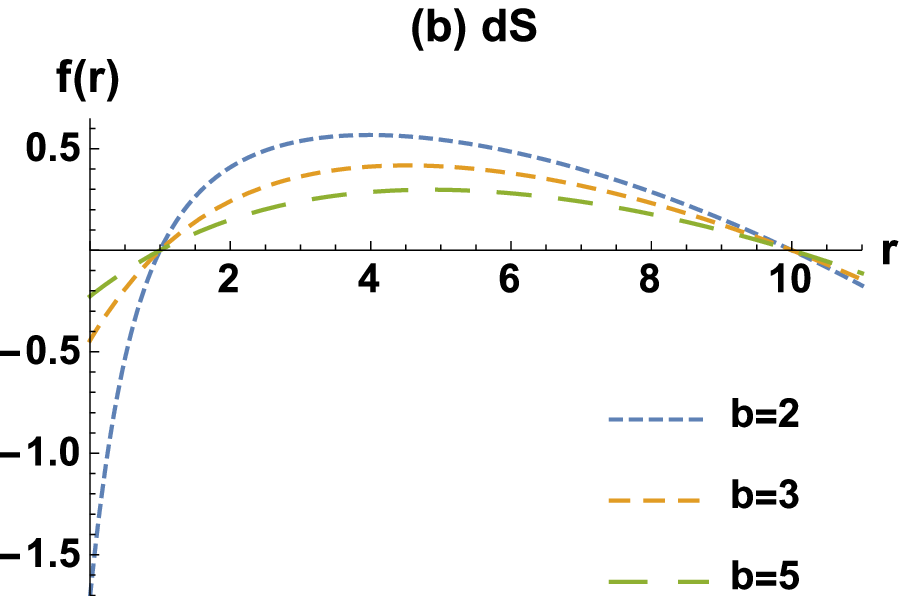}\includegraphics[height=4cm]{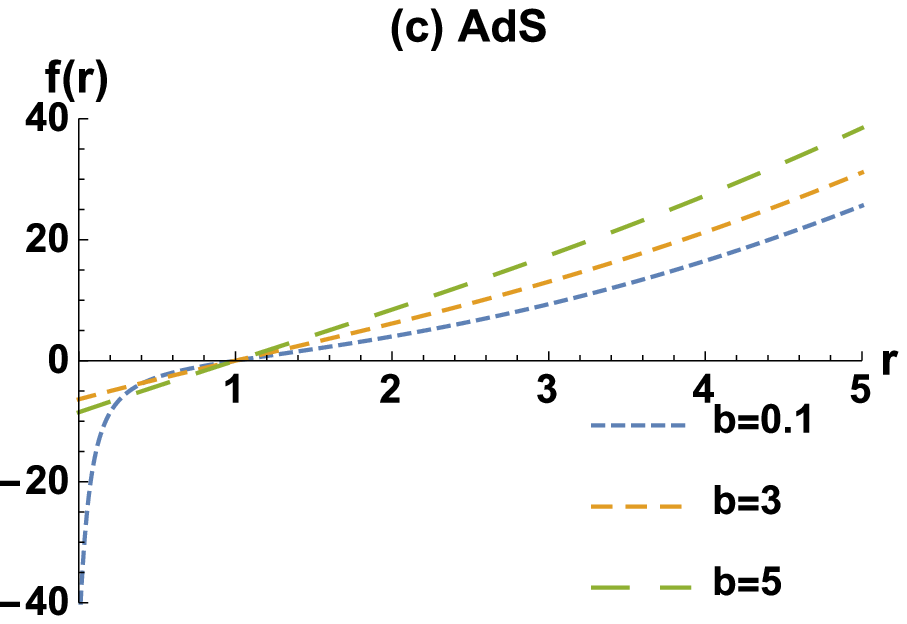}
\caption{The structures of the regular phantom BH's metric function $f(r)$ for different values of $b$: (a) asymptotically flat spacetime with $r_{\text{h}}=1$. (b) de Sitter spacetime with $r_{\text{h}}=1, r_{\text{c}}=10$ (c) anti-de Sitter spacetime with $r_{\text{h}}=1$.}\label{f}.
\end{figure}

Since the parameters $c$, $m$ can be expressed in terms of $b$, $r_{\text{h}}$ and $r_{\text{c}}$ (dS), the structures of the regular phantom BH spacetime are completely determined by $b$,$r_{\text{h}}$ and $r_{c}$(dS) (cf. Fig.\ref{f}). 
One can readily verify that all the spacetimes are indeed nonsingular even at $r=0$. 
As for any asymptotically flat spacetime, such flat regular phantom BH has a Schwarzschild-like structure.
However, its tendency of approaching flat spacetime at infinity becomes slower with increasing $b$. 
In the dS case, the spacetime is bounded by two horizons, i.e., $r_{\text{h}}<r<r_{\text{c}}$. 
There is a maximum for $f(r)$, and it decreases with increasing $b$. 
For the AdS phantom BH, $f(r)\rightarrow r^2$ when $r\rightarrow\infty$, and for larger $b$, the approach to the asymptotic solution becomes slower. 

\section{Gravitational quasinormal frequencies for regular phantom black holes}
\label{3}
As proposed by Regge-Wheeler, two important gravitational perturbations are of odd and even parity.
The perturbation gauge $h_{\mu\nu}$ for each type has its own definition. 
In this section, we choose the Regge-Wheeler-Zerilli gauge to discuss the master equation for each perturbation type. 
Here we take the magnetic quantum $M=0$ to make $\varphi$ disappear completely because all values of $M$ lead to the same radial equation \cite{wheeler}. 
Since the total number of equations in the even parity case is bigger than that in odd parity case, the derivation of the master equation for even parity perturbation is thus more complicated. 
Once the master equation is derived, the effective potential and the corresponding qusinormal modes can be obtained. 
We will first discuss the odd parity gravitational quasinormal frequencies in asymptotically flat, dS and AdS spacetimes, then study the case for even parity. 
In our work, we consider the metric perturbations not only of the Ricci curvature tensor and scalar curvature (the l.h.s. of the Einstein field equation), but also of the energy momentum tensor (the r.h.s. of the Einstein field equation). 
On the other hand, we will not consider the perturbations of the phantom scalar field.
This is because such perturbations can be canceled out through an appropriate choice of $V$ in the action (see Appendix for details).

In order to discuss gravitational perturbation, one may write down
\begin{equation}
g_{\mu\nu}=\bar{g}_{\mu\nu}+h_{\mu\nu}.
\label{eq:finalmetric}
\end{equation}
where the small perturbation $h_{\mu\nu}$ will be divided into odd and even modes in the subsequent sections.

\subsection{Master equation and quasinormal modes for odd parity perturbation}
The odd parity perturbation $h_{\mu\nu}$ has the form as\cite{wheeler}
\begin{equation}
h_{\mu\nu}=\left(
\begin{array}{cccc}
 0 & 0 & 0 & h_{0}(r,t) \\
 0 & 0 & 0 & h_{1}(r,t)\\
 0 & 0 & 0 & 0\\
 h_{0}(r,t) & h_{1}(r,t) & 0 & 0
\end{array}
\right)Q_{p}(\theta),
\label{eqh}
\end{equation}
where $Q_{p}(\theta)=\sin\theta dP_{l}(\cos\theta)/d\theta$ ($P_{l}(\cos\theta)$ is the Legendre function), which satisfies
\begin{equation}
Q_{p}''-\text{cot}(\theta)Q_{p}'(\theta)=-kQ_{p}(\theta),
\label{Qp}
\end{equation}
where $k$$=l(l+1)$,$l$ is the angular quantum number.

Then the separation of variables can be carried out by writing $h_{0}(r,t)=\exp(-i\omega t)h_{0}(r)$, $h_{1}(r,t)=\exp(-i\omega t)h_{1}(r)$.
By substituting Eq.(\ref{fr})$-$(\ref{Vr}), (\ref{eq:finalmetric})$-$(\ref{Qp}) into the field equation Eq.(\ref{fieldequation}) and only keeping the first order perturbation terms, we obtain the independent perturbation equations as follows:
\ba &&
\delta \widehat{G}_{13}=h_{0}'-\frac{2h_{0}p'}{p}-\frac{1}{\omega p^{2}}\left\{ih_{1}(-\omega^{2}p^{2}+f(-2+l+l^{2}+2pf'p'+p^{2}(2V(\phi)+f''))
\right.\nonumber\\ && \left.
+f^{2}p(\epsilon p\phi'^{2}+2p''))\right\}=0,
\label{Go13}
\ea
\begin{equation}
\delta \widehat{G}_{23}=\frac{i\omega h_{0}}{f^{2}}+\frac{h_{1}f'}{f}+h_{1}'=0.
\label{Go23}
\end{equation}
Eq.(\ref{Go23}) implies $h_{0}=-\frac{f^{2}}{i\omega}(\frac{h_{1}f'}{f}+h_{1}')$. Substituting $h_{0}$ into Eq.(\ref{Go13}), we get the master equation for odd parity perturbation
\ba &&
h''_{1}+(\frac{3f'}{f}-\frac{2p'}{p})h_{1}'+\frac{h_{1}}{f^{2}p^{2}}\left\{p^{2}(\omega^{2}+f'^{2})-f(-2+l+l^{2}+2p^{2}V(\phi)+4pf'p')
\right.\nonumber\\ && \left.
-f^{2}p(\epsilon p\phi'^{2}+2p'')\right\}=0.
\label{mastereq}
\ea
Finally, we renormalize $h_{1}$ by
\begin{eqnarray}
h_{1}(r)=B(r)\Phi(r);B(r)=\frac{p(r)}{f(r)}.
\label{redefine1}
\end{eqnarray}
By substituting Eq.(\ref{redefine1}) into the master equation, Eq.(\ref{mastereq}), and using a tortoise coordinate $r_{*}$, the Schr$\ddot{o}$dinger-type wave equation for this case can be expressed as
\begin{equation}
\frac{d^{2}\Phi}{dr_{*}^{2}}+(\omega^{2}-V_{o}(r))\Phi=0,
\label{masterodd}
\end{equation}
where the effective potential for odd parity perturbations $V_{o}(r)$ is
\begin{equation}
V_{o}(r)=\frac{f}{p^{2}}[-2+l+l^{2}+2fp'^{2}+p^{2}(2V(\phi)+\epsilon f\phi'^{2}+f'')+p(f'p'+fp'')].
\label{Vo}
\end{equation}

Eq.(\ref{Vo}) can be used to describe the effective potential of ``odd"-type perturbation $V_{o}$ in different spacetimes, and be utilized to discuss the relationship between the effective potential and model parameters such as the angular harmonic index $l$ and the parameter $b$.

\begin{figure}
\includegraphics[height=4cm]{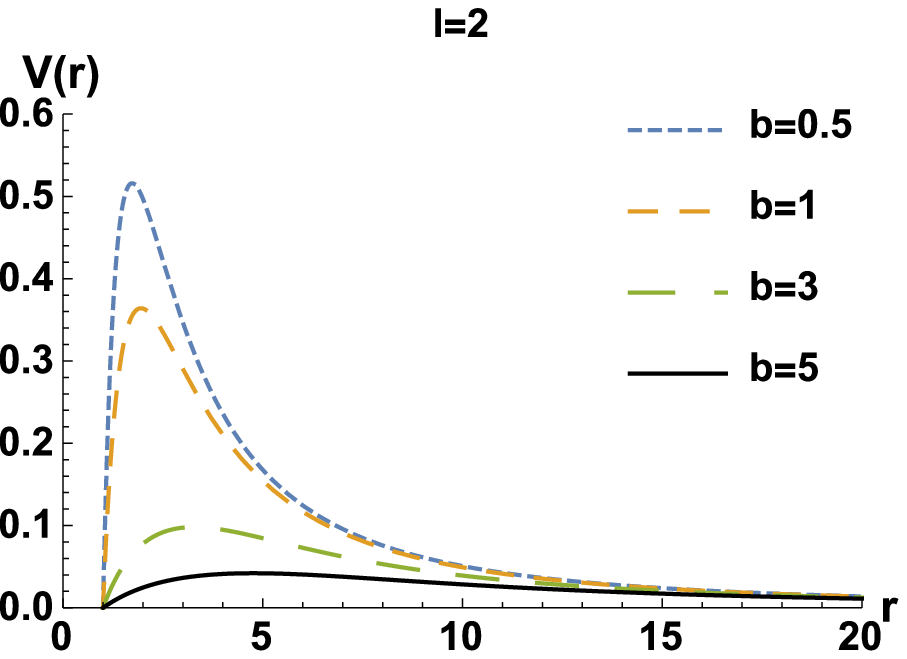}\includegraphics[height=4cm]{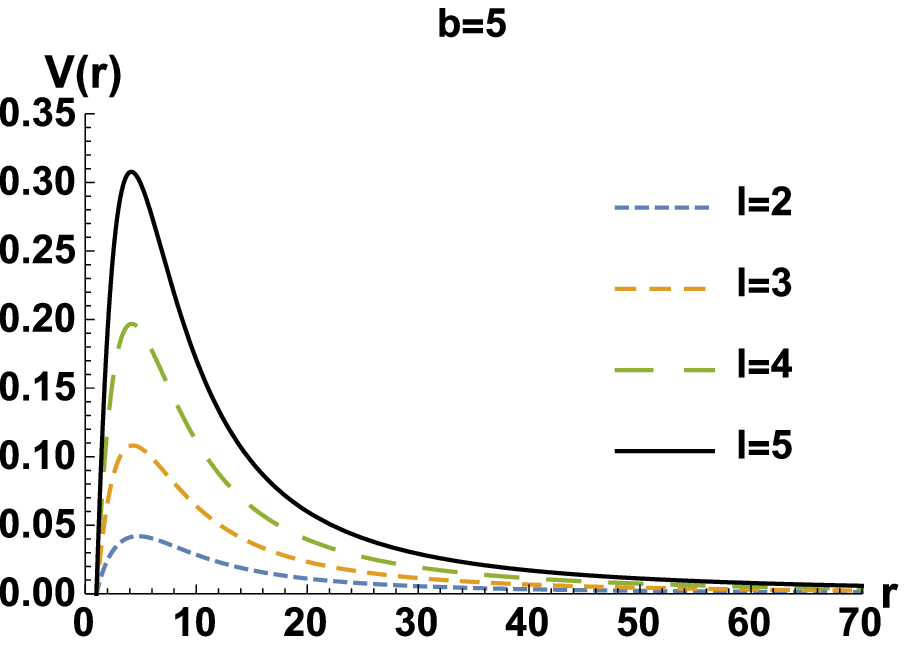}
\caption{The effective potential for odd parity gravitational perturbation in asymptotically flat spacetime, with $r_{\text{h}}=1, l=2$ (left), and with $r_{\text{h}}=1,b=5$ (right).}\label{Vflat}
\end{figure}

\begin{figure}[htbp]
\centerline{\includegraphics[height=4.5cm]{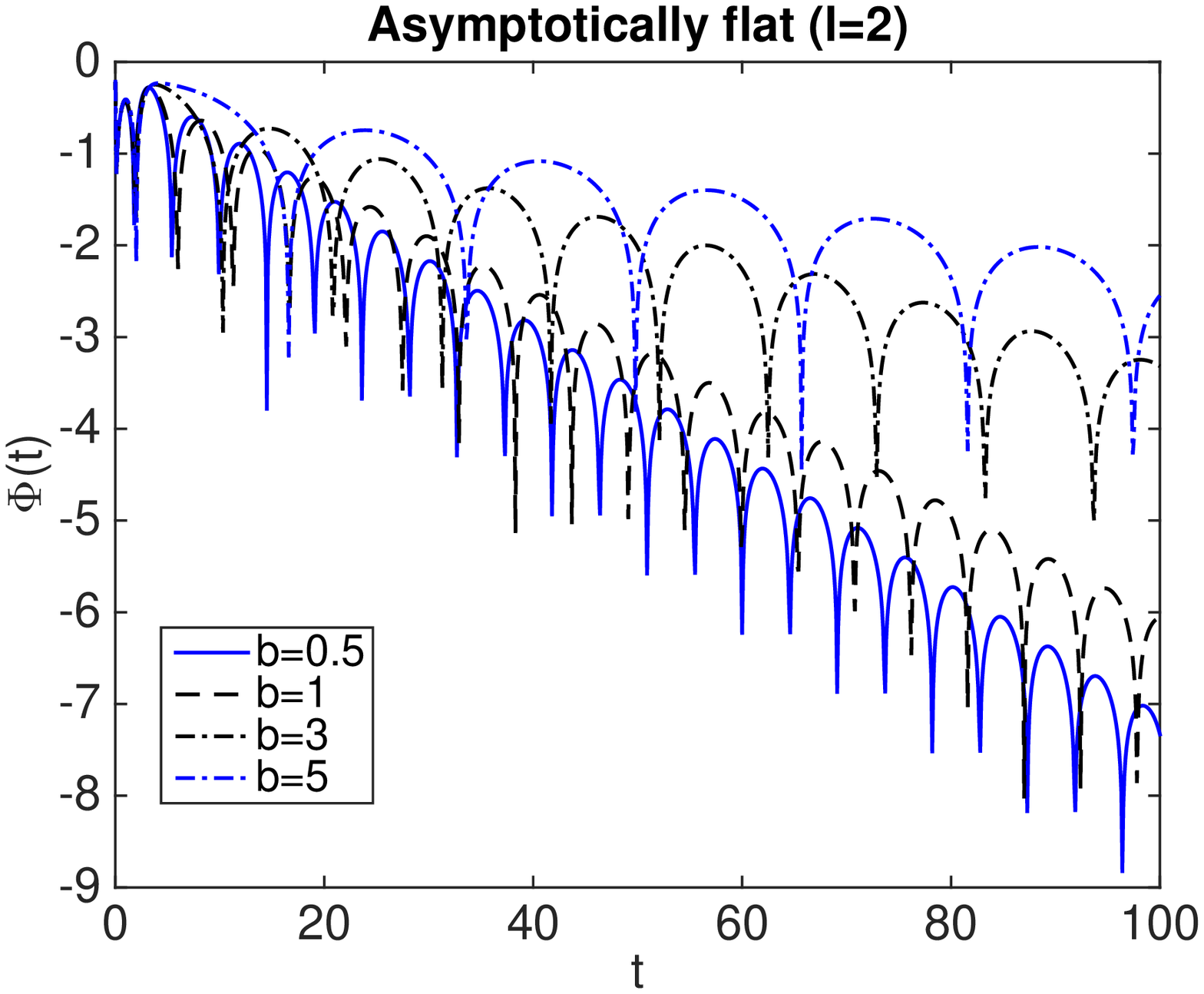}\includegraphics[height=4.5cm]{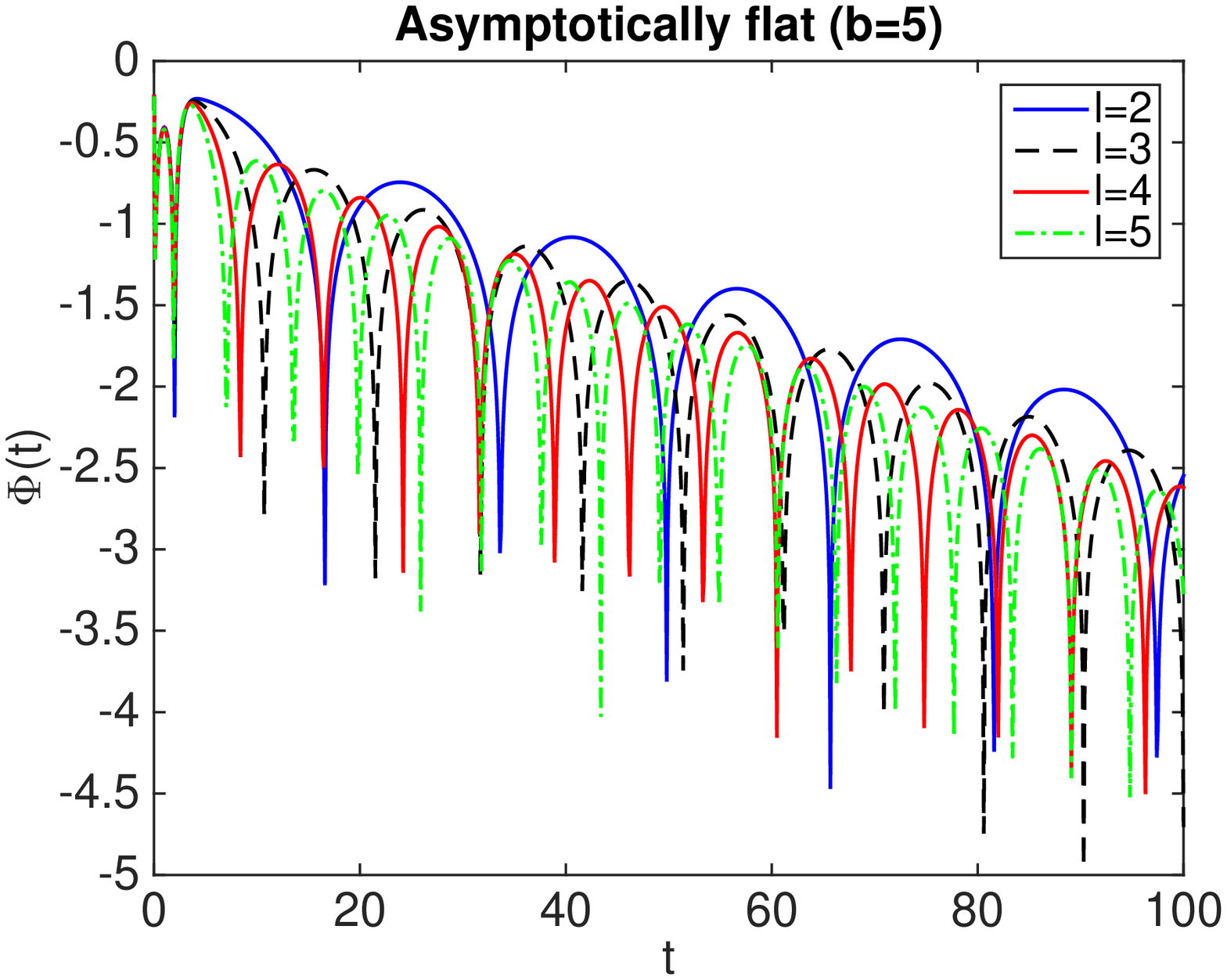}}
\caption{The tempoeral evolution of odd parity gravitational perturbation in asymptotically flat phantom BH, with $r_{\text{h}}=1, l=2$ (left), and with $r_{\text{h}}=1, b=5$ (right).}\label{finiteflat}
\end{figure}

\begin{figure}[htbp]
\centerline{\includegraphics[height=5.5cm]{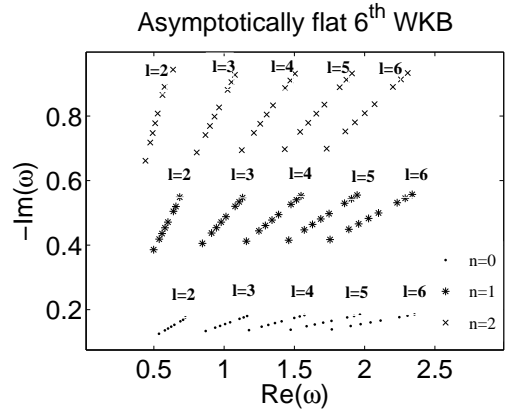}}
\caption{Calculated quasinormal frequencies of odd parity gravitational perturbation for $r_{\text{h}}=1$ in asymptotically flat spacetime. 
Each group of dots, from top-right to bottom-left, corresponds to QNM frequencies obtained by assuming different values of $b= 0.3, 0.4, 0.5, 0.7, 0.8, 0.9, 1.0, 1.2$, respectively.}\label{WKB}
\end{figure}

I) Fig.\ref{Vflat}, \ref{finiteflat} and \ref{WKB} show the potential functions, temporal evolution of the gravitational perturbation, and quasinormal frequency obtained by WKB method in asymptotically flat spacetime.

\begin{itemize}
\item The form of the effective potential as a function of $r$ for different values of $l$ and $b$ is shown in Fig.\ref{Vflat}. 
From the left plot, one sees that as $b$ increases, the shape of the effective potential becomes smoother.
The maximum of the effective potential decreases with increasing $b$ and the position of the peak shifts to the right. 
From the right plot, we see that with the increase of angular quantum number $l$, the maximum of the effective potential also increases. 
$V(r)$ is always found to be positive definite outside the event horizon, which indicates that the corresponding QNMs are likely to be stable.
\end{itemize}

\begin{itemize}

\item We adopt the finite difference method to analyze the stability of such BH. 
By applying the coordinate transformation $(t, r) \rightarrow ( \mu, \nu)$ with $\mu=t-r_{*}$, $\nu=t+r_{*}$ to Eq.(\ref{mastereq}), and integrating numerically using the finite difference method \cite{finit1,finit2,finit3,finit4}, we obtain the differential equation for $h_{1}(\mu,\nu)$. 
Fig.\ref{finiteflat} shows the stability of a regular phantom BH in asymptotically flat spacetime with $r_{\text{h}}=1$. 
The temporal evolution of each mode in Fig.\ref{finiteflat} corresponds to a corresponding case in Fig.\ref{Vflat}. 
Since $V(r)$ decreases significantly with the growth of $b$, the decay rate ($|\text{Im}(\omega)|$) becomes smaller and oscillation frequency (i.e., $\text{Re}(\omega)$) also drops; 
As $l$ increases, the values of $V(r)$ are raised correspondingly, so that the oscillation frequency (i.e., $\text{Re}(\omega)$) slightly increases together with the decay rate ($|\text{Im}(\omega)|$).
\end{itemize}

\begin{itemize}
\item We employ the WKB approximation \cite{WKB1,WKB2} to evaluate the quasinormal frequencies. The complex frequency $\omega$ is determined by\cite{WKB3}
\begin{equation}
\omega^{2}=\left[V_{0}+(-2V''_{0})^{1/2}P\right]-i\left(n+\frac{1}{2}\right)(-2V''_{0})^{1/2}(1+\Omega),~~(3^{\text{rd}} \text{order})
\end{equation}
\begin{equation}
i\frac{\omega^{2}-V_{0}}{\sqrt{-2V''_{0}}}-P-\Omega-P_{4}-P_{5}-P_{6}=n+\frac{1}{2},~~(6^{\text{th}} \text{order})
\end{equation}
where $V^{(n)}_{0}=\frac{d^{n}V}{dr^{n}_{*}}|_{r_{*}=r_{*}(r_{\text{h}})}, P, \Omega, P_{4}, P_{5}$, and $P_{6}$ are presented in\cite{WKB3,WKB4} .

By making use of Eq.(\ref{Vo}), we evaluate the QNM frequencies by employing the $6^{\rm{th}}$ order WKB method (see Fig.\ref{WKB}). 
Figure \ref{WKB} shows that the fundamental quasinormal modes ($n=0$) have the smallest imaginary parts, as the modes decay the slowest. 
As $n$ increases, the imaginary part of the corresponding quasinormal mode becomes bigger for given $l$ and $b$. 
For a given principal quantum number $n$, both the real and the imaginary parts of the frequency decrease with increasing $b$; on the other hand, the real part of the frequency increases significantly with the angular quantum number $b$, while the imaginary part also increases slightly with increasing $b$.
\end{itemize}

\begin{figure}[htbp]
\centerline{\includegraphics[height=4.5cm]{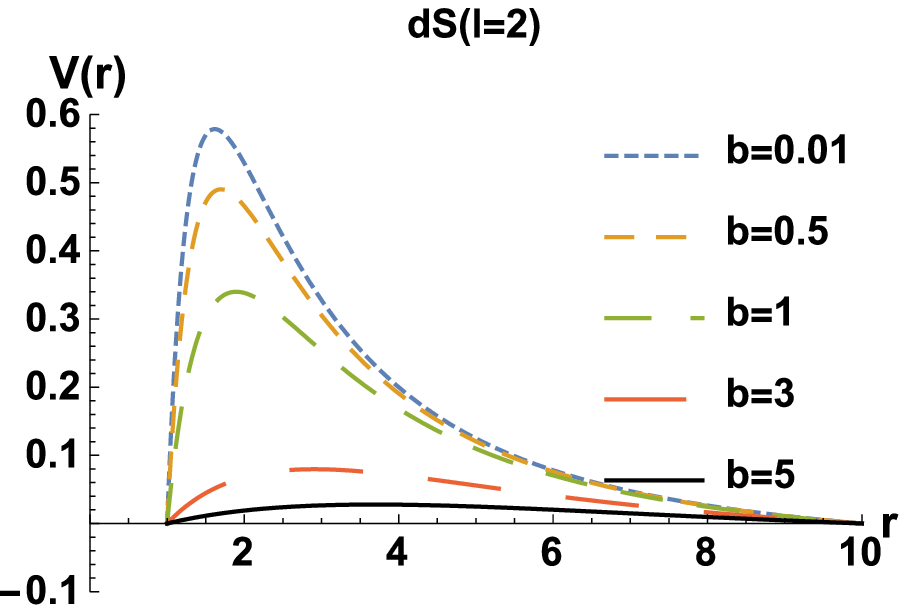}\includegraphics[height=4.5cm]{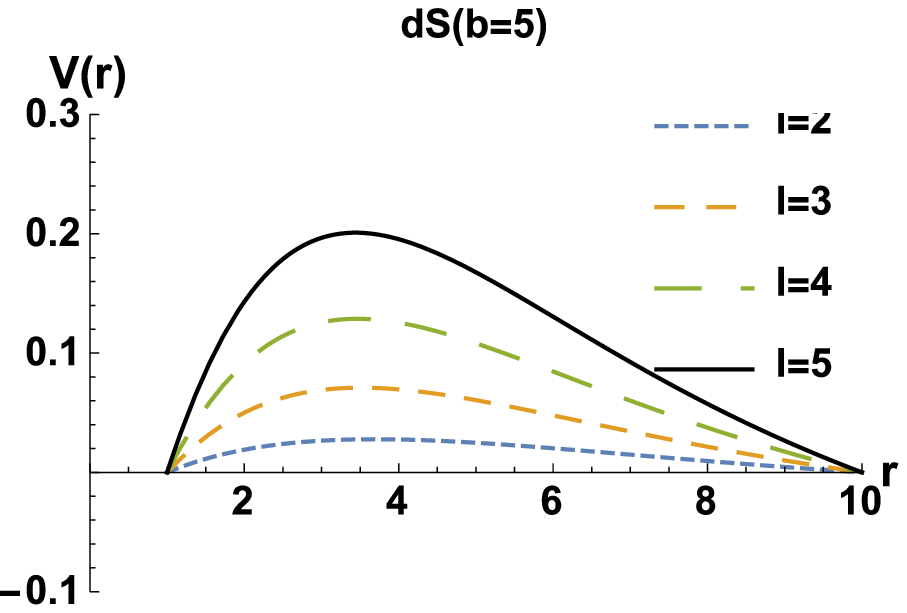}}
\caption{The effective potential for odd parity gravitational perturbation in dS spacetime with $r_{\text{h}}=1, r_{\text{c}}=10, l=2$ (left), and with $r_{\text{h}}=1, r_{\text{c}}=10, b=5$ (right).}\label{Vds}
\end{figure}

\begin{figure}[htbp]
\centerline{\includegraphics[height=4.5cm]{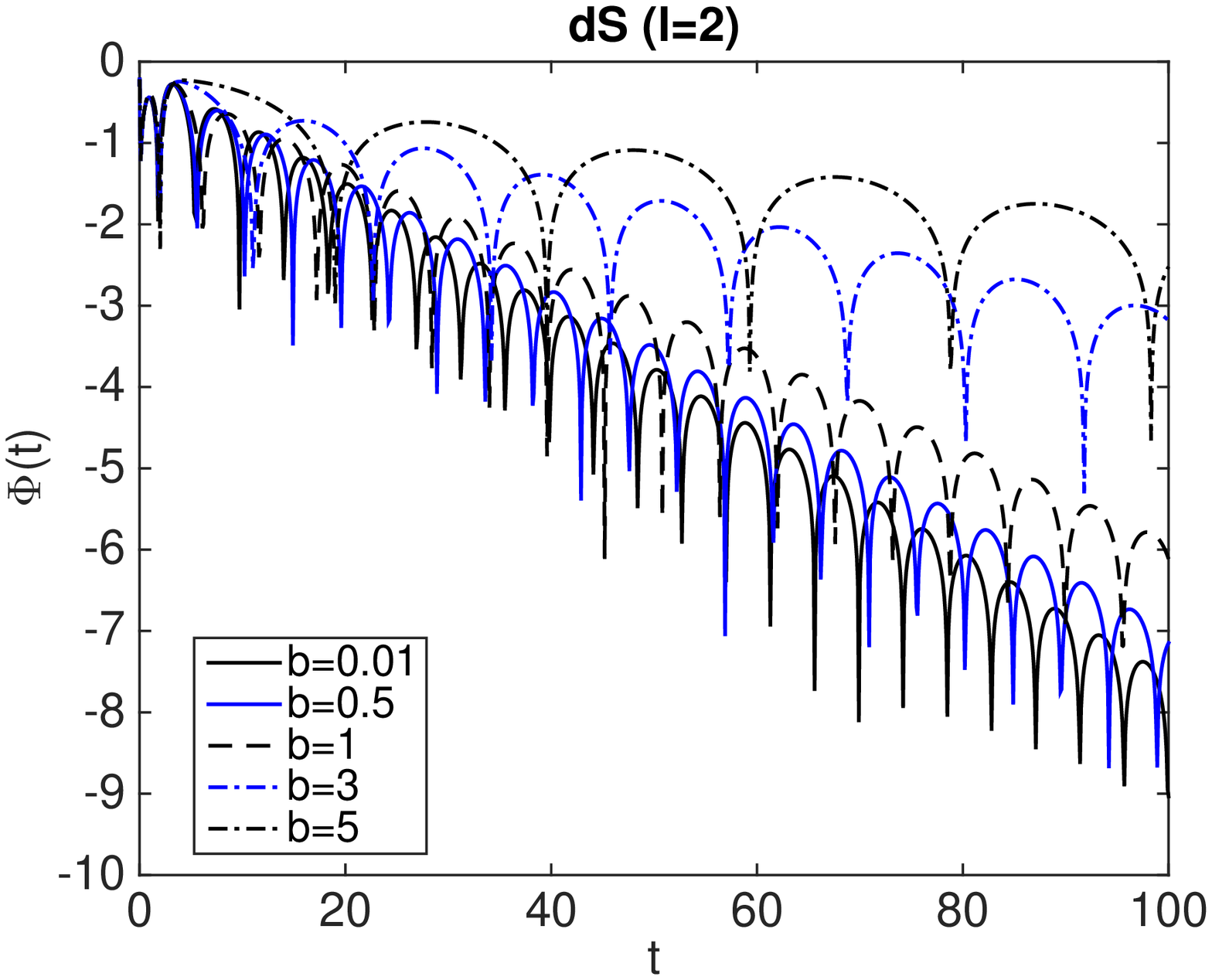}\includegraphics[height=4.5cm]{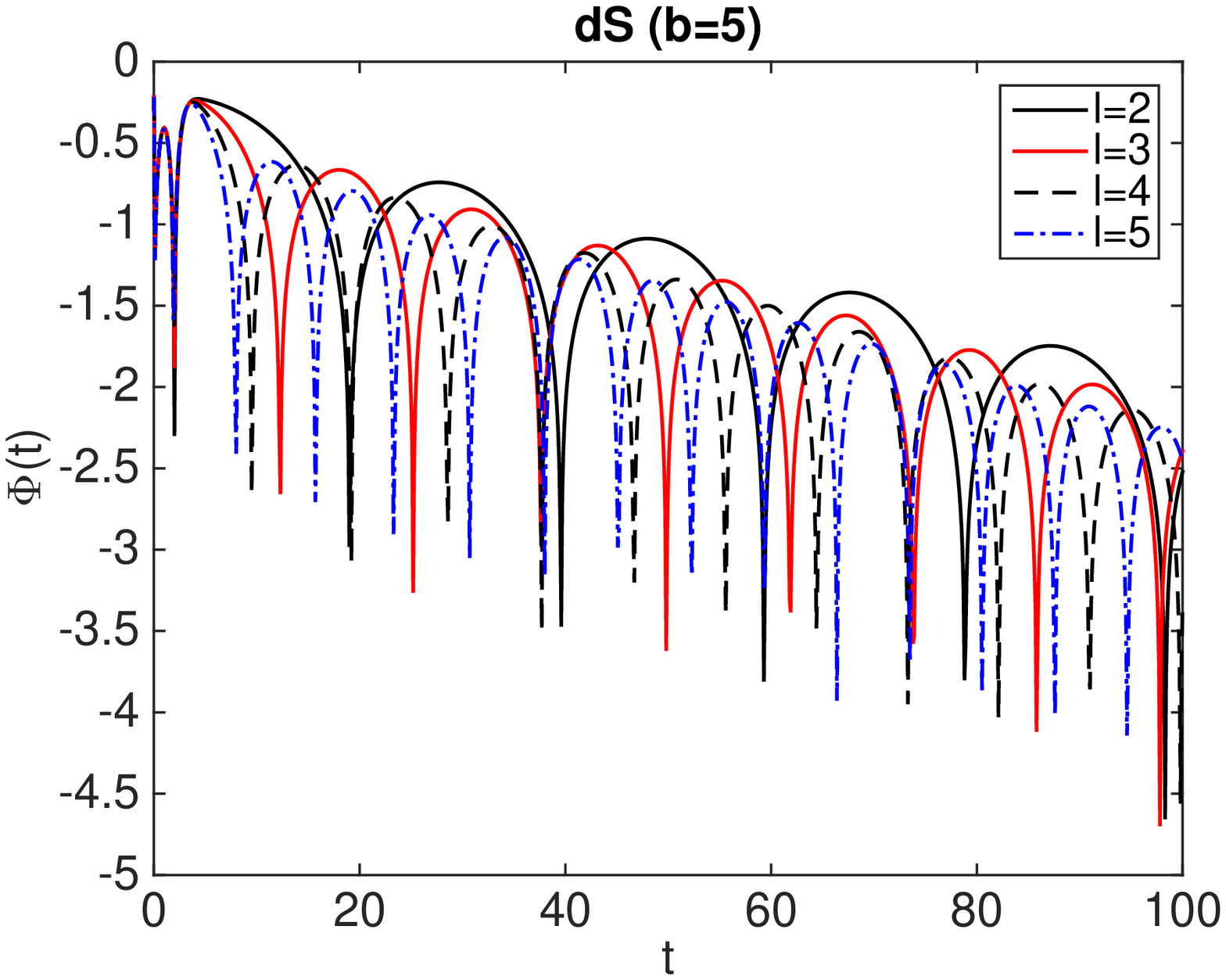}}
\caption{The temporal evolution of odd parity gravitational perturbation for dS phantom BH with $r_{\text{h}}=1, l=2$ (left), and with $r_{\text{h}}=1, b=5$ (right).}\label{finiteds}
\end{figure}

\begin{figure}[htbp]
\centerline{\includegraphics[height=5.5cm]{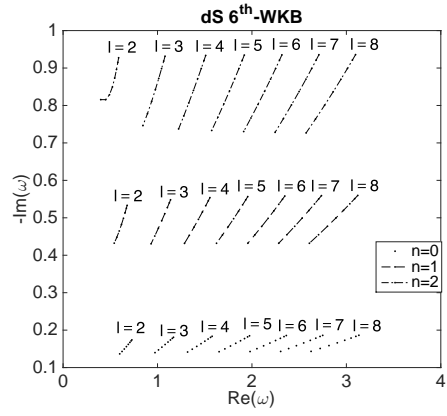}}
\caption{Calculated quasinormal frequencies of odd parity gravitational perturbation with $r_{\text{h}}=1, r_{\text{c}}=10$ in dS spacetime. 
Each group of dots, from top-right to bottom-left, corresponds to QNM frequencies obtained by assuming different values of $b=0.1,0.2, 0.3, 0.4, 0.5, 0.6, 0.7, 0.8$, respectively.}\label{WKBds}
\end{figure}

II) Fig. \ref{Vds},\ref{finiteds} and \ref{WKBds} show the potential function between the event horizon $r_{\text{h}}$ and cosmological horizon $r_{c}$, temporal evolution of the gravitational field, and quasinormal frequency obtained by WKB method in dS spacetime.
\begin{itemize}
\item The form of the effective potential in dS spacetime is similar to that in asymptotically flat spacetime. Fig.\ref{Vds} indicates that for given $l$ and $r$, the effective potential decreases with increasing $b$; and for given $b$ and $r$, it increases with increasing angular quantum number $l$. 
In the range $r_{\text{h}}<r<r_{c}$, $V(r)$ is also positive definite, which implies that the corresponding QNM is likely to be stable.
\end{itemize}

\begin{figure}[htbp]
\centerline{\includegraphics[height=4.5cm]{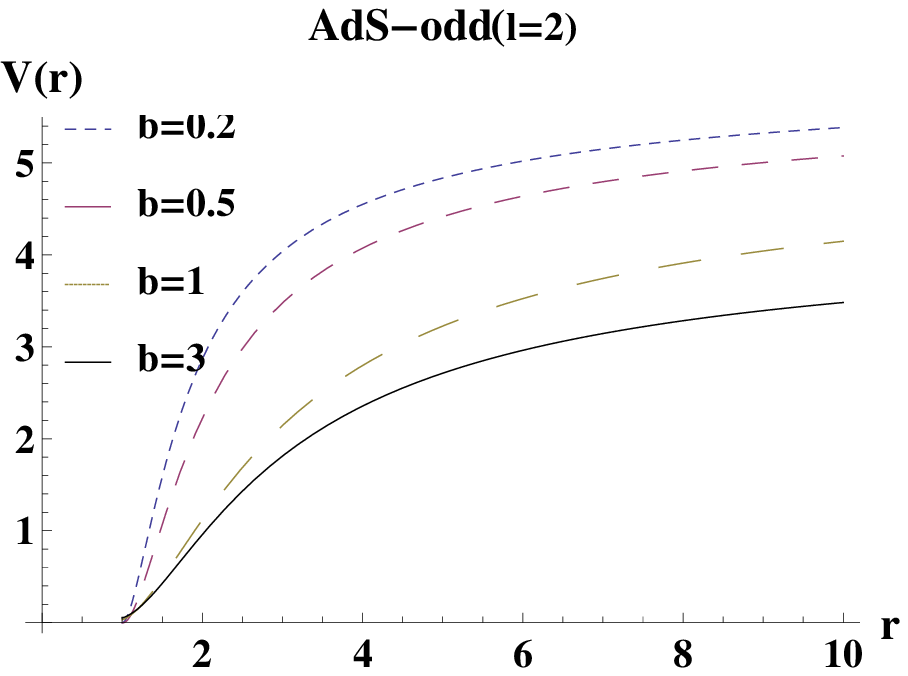}\includegraphics[height=4.5cm]{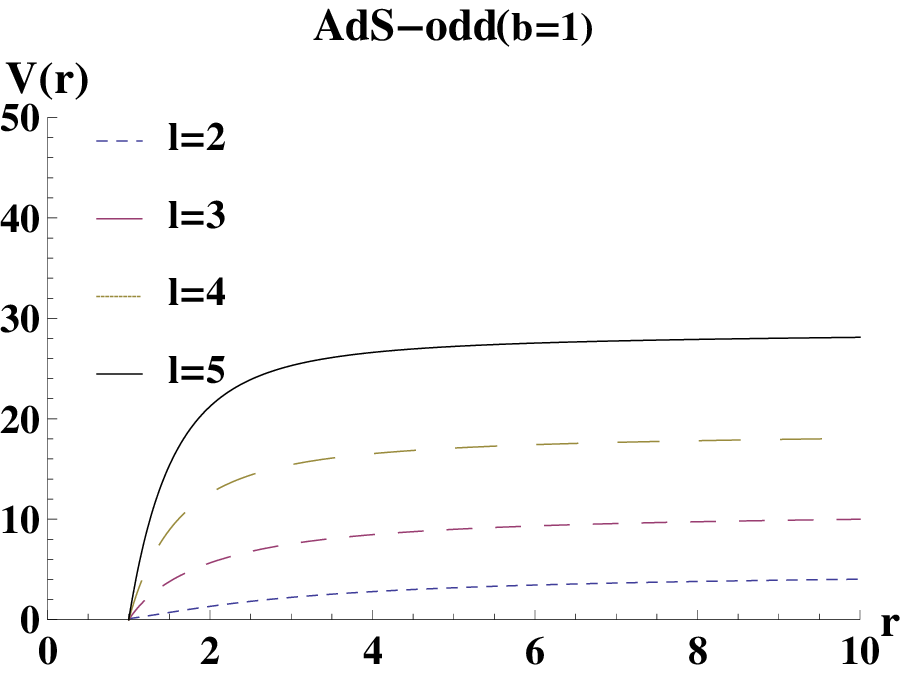}}
\caption{The effective potential for odd parity gravitational perturbation in AdS spacetime with $r_{\text{h}}=1, l=2$ (left), and with $r_{\text{h}}=1, b=1$ (right).}\label{Vads}
\end{figure}

\begin{figure}[htbp]
\centerline{\includegraphics[height=4.5cm]{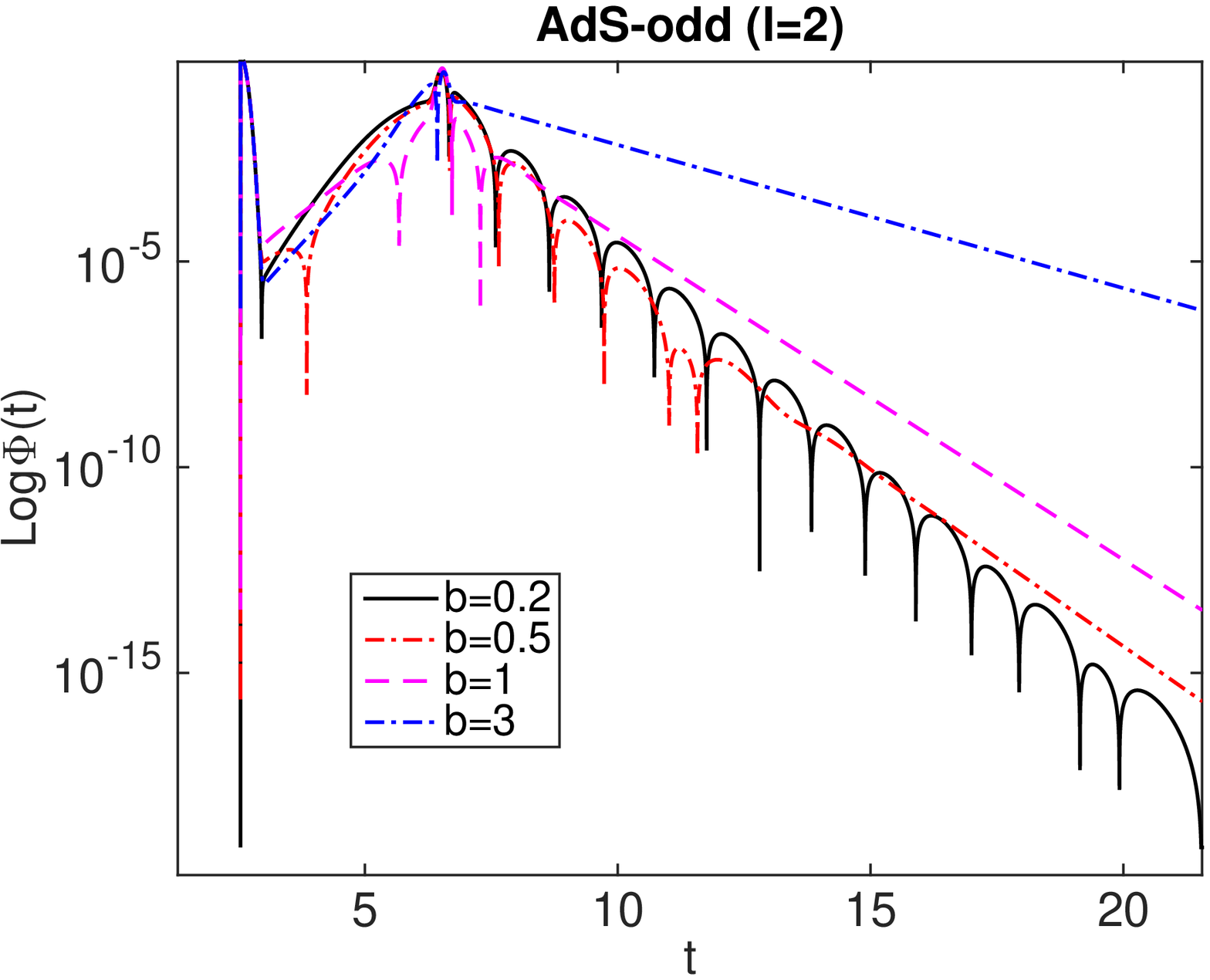}\includegraphics[height=4.5cm]{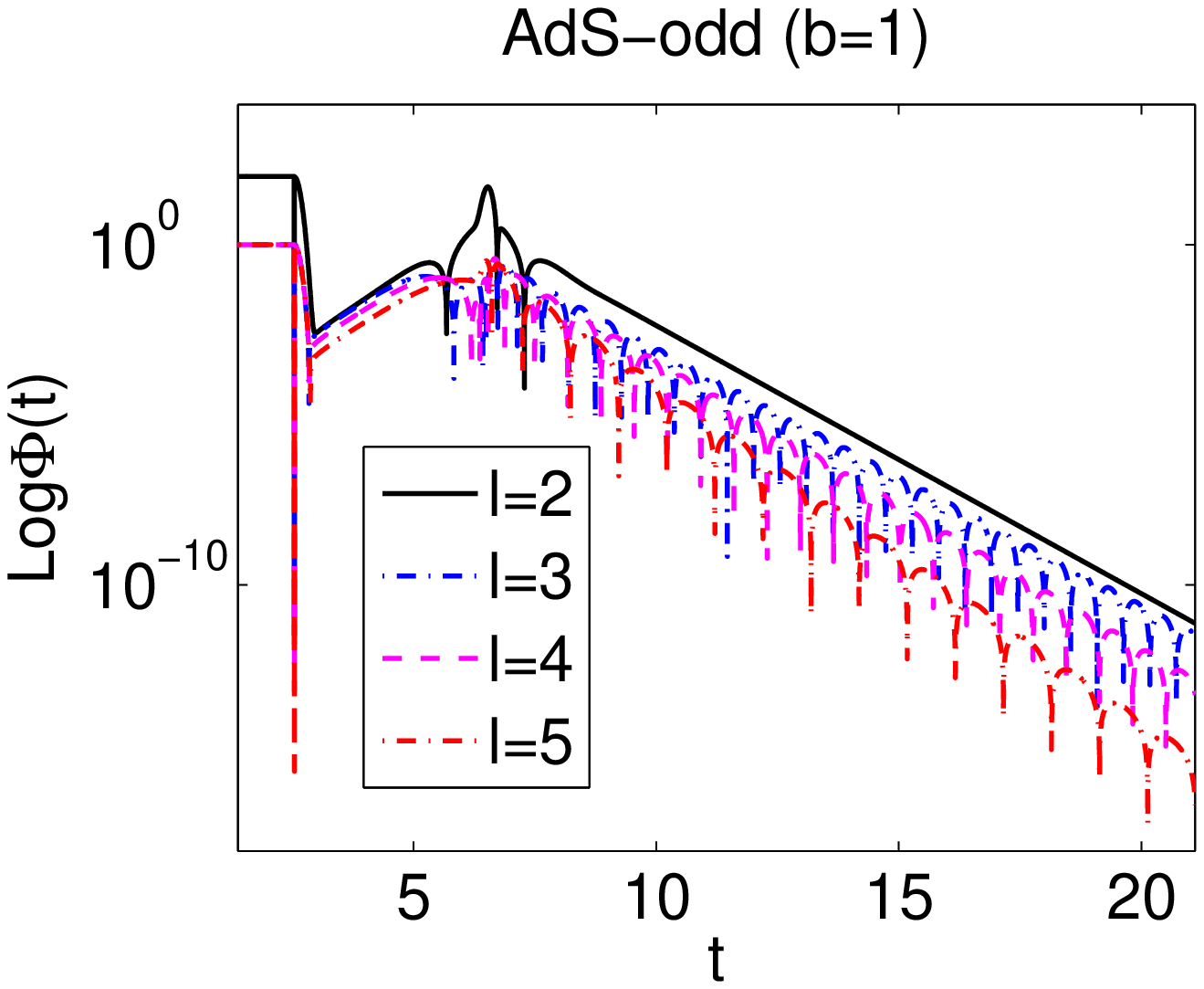}}
\caption{The temporal evolution of odd parity gravitational perturbation in AdS phantom BH spacetime with $r_{\text{h}}=1, l=2$ (left), and with $r_{\text{h}}=1, b=1$ (right).}\label{finiteads}
\end{figure}

\begin{itemize}
\item Fig.\ref{finiteds} studies the stability of a regular phantom with $r_{\text{h}}=1, r_{\text{c}}=10$. 
Similar to the case in asymptotically flat spacetime, it is found with increasing $b$, the oscillation frequency (i.e., $\text{Re}(\omega)$) decreases, while the decay rate ($|\text{Im}(\omega)|$) becomes smaller. 
Therefore, for smaller value of $b$, the BH returns to its stable state more quickly as small perturbation dies out faster.
\end{itemize}
\begin{itemize}
\item In accordance with the results of finite difference method, the frequencies presented in Fig.\ref{WKBds} by WKB method also show that for given $n$ and $l$, the decay rate of perturbation (i.e., $|\text{Im}(\omega)|$) decreases with increasing $b$. 
Moreover, it illustrates that for a given $b$, the fundamental mode can be found at $n=0, l=2$. 
Due to the smallness of the real parts of the frequencies, these modes of oscillation persist for longer time.
\end{itemize}
III) Fig.\ref{Vads} and \ref{finiteads} show the potential function beyond the event horizon $r>r_{\text{h}}$ and the temporal evolution of the gravitational perturbation in AdS spacetime.
\begin{itemize}
\item The form of the effective potential in AdS space time is quite different from that in asymptotically flat and dS spacetime. 
Fig.\ref{Vads} indicates that the value of $V(r)$ is divergent as $r\rightarrow \infty$. 
The effects of $b$ and $l$ on the effective potential are similar to previous cases, for a given $r$, smaller $b$ or larger $l$ leads to bigger $V$.
\end{itemize}

\begin{itemize}
\item Fig.\ref{finiteads} studies the stability of AdS regular phantom BH spacetime with $r_{\text{h}}=1$. 
The results are consistent with the above calculated potential function. 
It is again inferred that the fundamental mode of gravitational perturbation in odd parity occurs for $l=2$ and larger $b$, since such kind of QNM will take a longer time to be stable.
\end{itemize}

\subsection{Master equation and quasinormal modes for even parity perturbation}
Another canonical form for the gravitational perturbations is of even parity. After applying the separation of variables, it can be expressed as \cite{wheeler,Gq4D1}
\begin{equation}
h_{\mu\nu}=\exp{(-i\omega t)}\left(
\begin{array}{cccc}
H_{0}(r)f(r) & H_{1}(r) & 0 & 0 \\
H_{1}(r) & H_{2}(r)/f(r) & 0 & 0\\
 0 & 0 & p(r)^{2}K(r) & 0\\
 0 & 0 & 0 & p(r)^{2}K(r)\sin^{2}\theta
\end{array}
\right)P_{l}(cos\theta),
\label{evenper}
\end{equation}
where $H_{0}(r), H_{1}(r), H_{2}(r), K(r)$ are unknown functions for the even parity perturbation. It is noted that these functions are not independent.

Now we derive the first order perturbation equations by substituting Eq.(\ref{fr})$-$(\ref{Vr}), (\ref{eq:finalmetric}) and (\ref{evenper}) into Eq.(\ref{fieldequation}), and find the relationships among $H_{0}(r), H_{1}(r), H_{2}(r), K(r)$.
\begin{eqnarray}
\delta \widehat{G}_{34}=0\Leftrightarrow H_{0}=H_{2}
\label{G34}
\end{eqnarray}
\begin{eqnarray}
\delta \widehat{G}_{12}=0\Leftrightarrow\frac{\chi_{1}(r)}{i\omega}H_{1}-\frac{p'}{p}H_{0}+K' +
 (-\frac{f'}{2 f} + \frac{p'}{p})K =0,
 \label{G12}
 \end{eqnarray}
\begin{eqnarray}
\delta \widehat{G}_{11}=0\Leftrightarrow-\chi_{0}(r)H_{0}-\frac{p'H'_{0}}{p}-\frac{(-2+l+l^{2})K}{2fp^{2}}+(\frac{f'}{2f}+\frac{3p'}{p})K'+K''=0,
 \label{G11}
 \end{eqnarray}

\begin{eqnarray}
\delta \widehat{G}_{23}=0\Leftrightarrow-\frac{i\omega H_{1}}{f}-\frac{f'H_{0}}{f}-H'_{0}+K'=0,
 \label{G23}
 \end{eqnarray}
\begin{eqnarray}
\delta \widehat{G}_{22}=0\Leftrightarrow \frac{2i\omega H_{1}}{f}+H'_{0}-(1+\frac{f'p}{2fp'})K'+\frac{(-2+l+l^{2})f-2\omega^{2}p^{2}}{2f^{2}pp'}K-\frac{-2+l+l^{2}+2p^{2}V(\phi)}{2fpp'}H_{0}=0,
 \label{G22}
 \end{eqnarray}
where
\begin{equation}
\chi_{1}(r)=\frac{-2+l+l^{2}+2fp'^{2}+p^{2}(2V(\phi)+\epsilon f\phi'^{2})+2p(f'p'+2fp'')}{2p^{2}},
\end{equation}
and
\begin{equation}
\chi_{0}(r)=\frac{-2+l+l^{2}+4fp'^{2}+2p^{2}(V(\phi)+\epsilon f\phi'^{2})+4p(f'p'+2fp'')}{2fp^{2}}.
\end{equation}
Solving the Eq.(\ref{G12}), $H_{0}$ can be expressed as
\begin{equation}
H_{0}=p\left\{\frac{\chi_{1}(r)}{i\omega}H_{1}+K'-(\frac{f'}{2f}-\frac{p'}{p})K\right\}/p'.
\label{H0a}
\end{equation}
We define a function $\Psi(r)$ satisfying
\begin{equation}
H_{1}=-i\omega\frac{r}{f}(\Psi+K).
\label{H1a}
\end{equation}
By observing Eqs.(\ref{G34}), (\ref{H0a}) and (\ref{H1a}), it turns out that we need to express $K$ by $\Psi$ in order to express all perturbation functions $H_{0}(r)$, $H_{1}(r)$, $H_{2}(r)$ and $K(r)$ in terms of $\Psi$. 
This can be achieved by substituting Eqs.(\ref{H0a}) and (\ref{H1a}) into Eqs.(\ref{G23}),({\ref{G22}}), and evaluating the subtraction $(\ref{G23})-({\ref{G22}})$, and one eventually obtains following expression
\begin{equation}
\frac{(-2+l+l^{2}+2p^{2}V(\phi)+3pf'p')K'}{2fp'^{2}}-(\frac{r\omega^{2}}{f^{2}}+\chi_{2}(r))\Psi-(\frac{r p'(r)-p(r)}{f(r)^2 p'(r)}\omega ^2+\chi_{3}(r))K=0,
\end{equation}
where
\begin{equation}
\chi_{2}(r)=\frac{r\left(-2+l+l^2+2 p^2 V(\phi)+2pf'p'\right)}{2 f^2p'^2}\chi_{1}(r),
\end{equation}
 and
\ba &&
 \chi_{3}(r)=\frac{1}{4 f^2 p'^2}(f' \left(-4 f p'^2+l^2+l+2 p^2 V(\phi)-2\right)
\nonumber\\ &&
 +2 r\chi_{1} \left(2pf'p'+l^2+l+2p^2V(\phi)-2\right)+2pf'^2 p'-4fpp' V(\phi)). \label{chi3}
  \ea
  Substituting Eq.(\ref{H0a})$-$(\ref{chi3}) into Eq.(\ref{G11}), $K$ can be solved as
  \begin{equation}
 K(r)=A_{1}(r)\Psi'(r)+A_{2}(r)\Psi(r),
 \label{KHa}
 \end{equation}
 where $A_{1}(r)$, $A_{2}(r)$ are
\ba &&
 A_{1}(r)=-(2 r f p^2 \chi_{1} p' \left(l^2+l+2 p^2 V(\phi)+3 p f' p'-2\right))\left\{4 p \chi_{3} p'^2 \left(p'' p+p'^2-p^2 \chi_{0}\right) f^3
\right. \nonumber\\ &&\left.
 -2 p^2 p' \left(\chi_{0} \left(l^2+l+2 p^2 V(\phi)+3 p f' p'-2\right)-2 \chi_{3} \left(r \chi_{1}+f'\right) p'^2\right) f^2+\left(2 V(\phi) \chi_{0} \left(2 r \chi_{1}+f'\right) p^5
 \right. \right. \nonumber\\ && \left.\left.
 +\left(\chi_{0} \left(4 \omega ^2+3 f'^2+6 r \chi_{1} f'\right) p'+2 V(\phi) \left(p' \left(2 r \chi'_{1}+f''\right)-f' p''+2 \chi_{1} \left(p'-r p''\right)\right)\right) p^4+\left(\chi_{0} \left(-4 r \omega ^2 p'^2
\right.\right.\right.\right. \nonumber\\ && \left.\left.\left.\left.
 +2 \left(l^2+l-2\right) r \chi_{1}+\left(l^2+l-2\right) f'\right)+p' \left(-4 p'' \omega ^2+2 V(\phi) \left(2 r \chi_{1}+f'\right) p'+6 r f' p' \chi'_{1}+3 f' p' f''-3 f'^2 p''
\right.\right.\right.\right.\nonumber\\ &&\left.\left.\left.\left.
 +6 \chi_{1} f' \left(p'-r p''\right)\right)\right) p^3+\left(-4 \omega ^2 p'^3+3 f'^2 p'^3+4 r \omega ^2 p'' p'^2-2 \left(l^2+l-2\right) V(\phi) p'+2 l^2 r \chi'_{1} p'+2 l r \chi'_{1} p'-4 r \chi'_{1} p'
\right.\right.\right.\nonumber\\ &&\left.\left.\left.
 +l^2 f'' p'+l f'' p'-2 f'' p'-l^2 f' p''-l f' p''+2 f' p''+2 \chi_{1} \left(3 r f' p'^3+\left(l^2+l-2\right) p'-\left(l^2+l-2\right) r p''\right)\right) p^2
\right.\right.\nonumber\\ &&\left.\left.
 +2 p'^2 \left(2 r \omega ^2 p'^2+\left(l^2+l-2\right) r \chi_{1}-\left(l^2+l-2\right) f'\right) p-\left(l^2+l-2\right)^2 p'\right) f-p^2 p' \left(2 r \chi_{1} \left(2 p' \left(p-r p'\right) \omega ^2
\right.\right.\right.\nonumber\\ &&\left.\left.\left.
 +\left(l^2+l+2 p^2 V(\phi)-2\right) f'+3 p f'^2 p'\right)+f' \left(4 p' \left(p-r p'\right) \omega ^2+\left(l^2+l+2 p^2 V(\phi)-2\right) f'+3 p f'^2 p'\right)\right)\right\}^{-1}.
 \label{A1}
 \ea
 \ba &&
A_{2}(r)= -(2 p \left(2 \chi_{2} p'^2 \left(p'' p+p'^2-p^2 \chi_{0}\right) f^3+2 p \chi_{2} \left(r \chi_{1}+f'\right) p'^3 f^2+\left(2 r V(\phi) \chi_{0} \chi_{1} p^4+\left(3 r \chi_{0} \chi_{1} f' p'
\right.\right.\right.\nonumber\\ &&\left.\left.\left.
+2 V(\phi) \left(r p' \chi'_{1}+\chi_{1} \left(p'-r p''\right)\right)\right) p^3+\left(r \chi_{0} \left(\left(l^2+l-2\right) \chi_{1}-2 \omega ^2 p'^2\right)+p' \left(2 r V(\phi) \chi_{1} p'+3 f' \left(r p' \chi'_{1}
\right.\right.\right.\right.\right.\nonumber\\ &&\left.\left.\left.\left.\left.
+\chi_{1} \left(p'-r p''\right)\right)\right)\right) p^2+\left(r p' \left(2 p' p'' \omega ^2+\left(l^2+l-2\right) \chi'_{1}\right)+\chi_{1} \left(3 r f' p'^3+\left(l^2+l-2\right) p'-\left(l^2+l-2\right) r p''\right)\right) p
\right.\right.\nonumber\\ &&\left.\left.
+r p'^2 \left(2 \omega ^2 p'^2+\left(l^2+l-2\right) \chi_{1}\right)\right) f+r p p' \left(2 \omega ^2 f' p'^2-\chi_{1} \left(3 p p' f'^2+\left(l^2+l+2 p^2 V(\phi)-2\right) f'-2 r \omega ^2 p'^2\right)\right)\right))
\nonumber\\ &&
\left\{4 p \chi_{3} p'^2 \left(p'' p+p'^2-p^2 \chi_{0}\right) f^3-2 p^2 p' \left(\chi_{0} \left(l^2+l+2 p^2 V(\phi)+3 p f' p'-2\right)-2 \chi_{3} \left(r \chi_{1}+f'\right) p'^2\right) f^2
\right.\nonumber\\ &&\left.
+\left(2 V(\phi) \chi_{0} \left(2 r \chi_{1}+f'\right) p^5+\left(\chi_{0} \left(4 \omega ^2+3 f'^2+6 r \chi_{1} f'\right) p'+2 V(\phi) \left(p' \left(2 r \chi'_{1}+f''\right)-f' p''+2 \chi_{1} \left(p'-r p''\right)\right)\right) p^4
\right.\right.\nonumber\\ &&\left.\left.
+\left(\chi_{0} \left(-4 r \omega ^2 p'^2+2 \left(l^2+l-2\right) r \chi_{1}+\left(l^2+l-2\right) f'\right)+p' \left(-4 p'' \omega ^2+2 V(\phi) \left(2 r \chi_{1}+f'\right) p'+6 r f' p' \chi'_{1}
\right.\right.\right.\right.\nonumber\\ &&\left.\left.\left.\left.
+3 f' p' f''-3 f'^2 p''+6 \chi_{1} f' \left(p'-r p''\right)\right)\right) p^3+\left(-4 \omega ^2 p'^3+3 f'^2 p'^3+4 r \omega ^2 p'' p'^2-2 \left(l^2+l-2\right) V(\phi) p'+2 l^2 r \chi'_{1} p'
\right.\right.\right.\nonumber\\ &&\left.\left.\left.
+2 l r \chi'_{1} p'-4 r \chi'_{1} p'+l^2 f'' p'+l f'' p'-2 f'' p'-l^2 f' p''-l f' p''+2 f' p''+2 \chi_{1} \left(3 r f' p'^3+\left(l^2+l-2\right) p'
\right.\right.\right.\right.\nonumber\\ &&\left.\left.\left.\left.
-\left(l^2+l-2\right) r p''\right)\right) p^2+2 p'^2 \left(2 r \omega ^2 p'^2+\left(l^2+l-2\right) r \chi_{1}-\left(l^2+l-2\right) f'\right) p-\left(l^2+l-2\right)^2 p'\right) f
\right.\nonumber\\ &&\left.
-p^2 p' \left(2 r \chi_{1} \left(2 p' \left(p-r p'\right) \omega ^2+\left(l^2+l+2 p^2 V(\phi)-2\right) f'+3 p f'^2 p'\right)+f' \left(4 p' \left(p-r p'\right) \omega ^2
\right.\right.\right.\nonumber\\ &&\left.\left.\left.
+\left(l^2+l+2 p^2 V(\phi)-2\right) f'+3 p f'^2 p'\right)\right)\right\}^{-1}.
 \label{A2}
 \ea
\begin{figure}[htbp]
\centerline{\includegraphics[height=4cm]{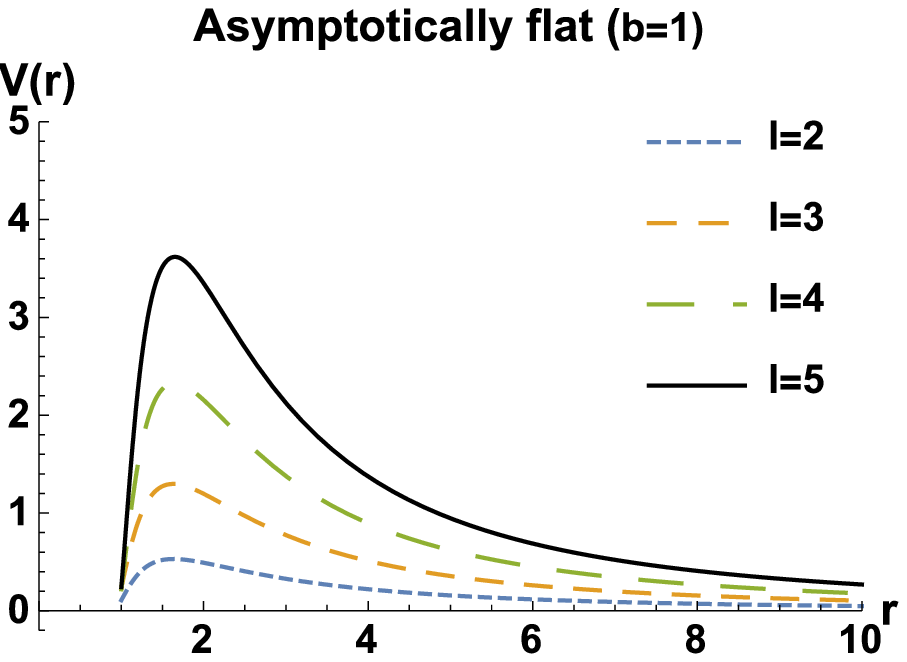}\includegraphics[height=4cm]{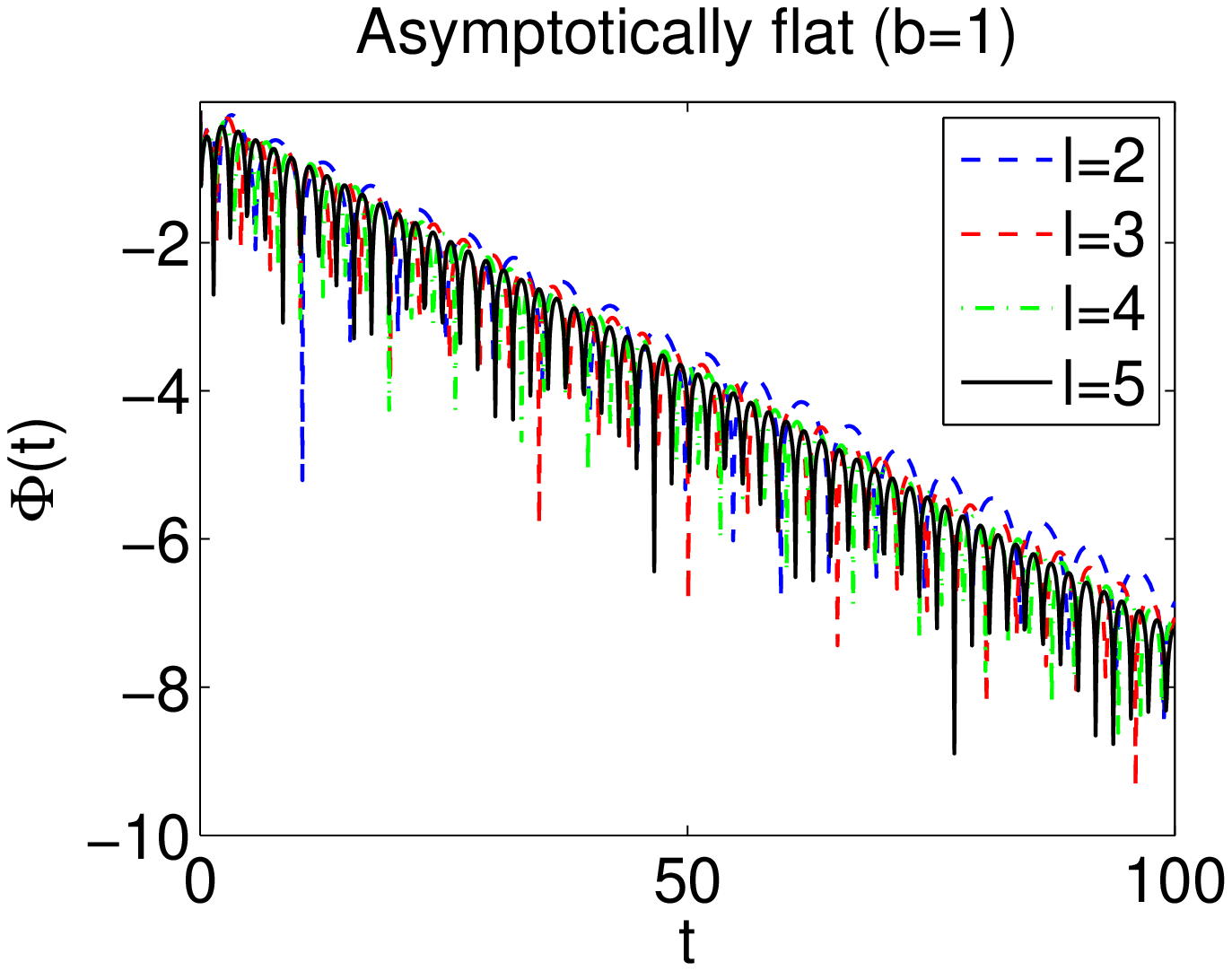}}
\centerline{\includegraphics[height=4cm]{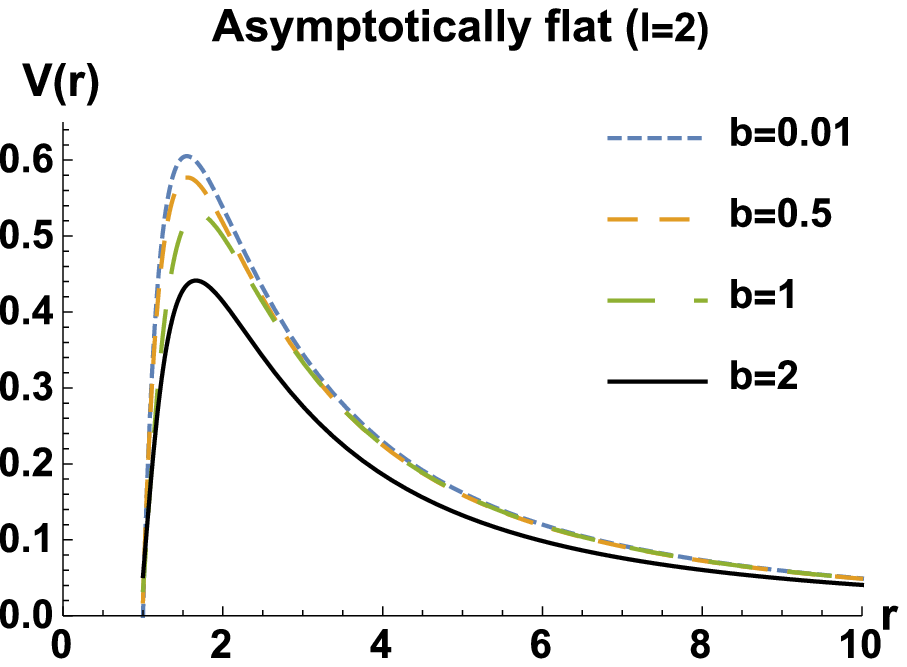}\includegraphics[height=4cm]{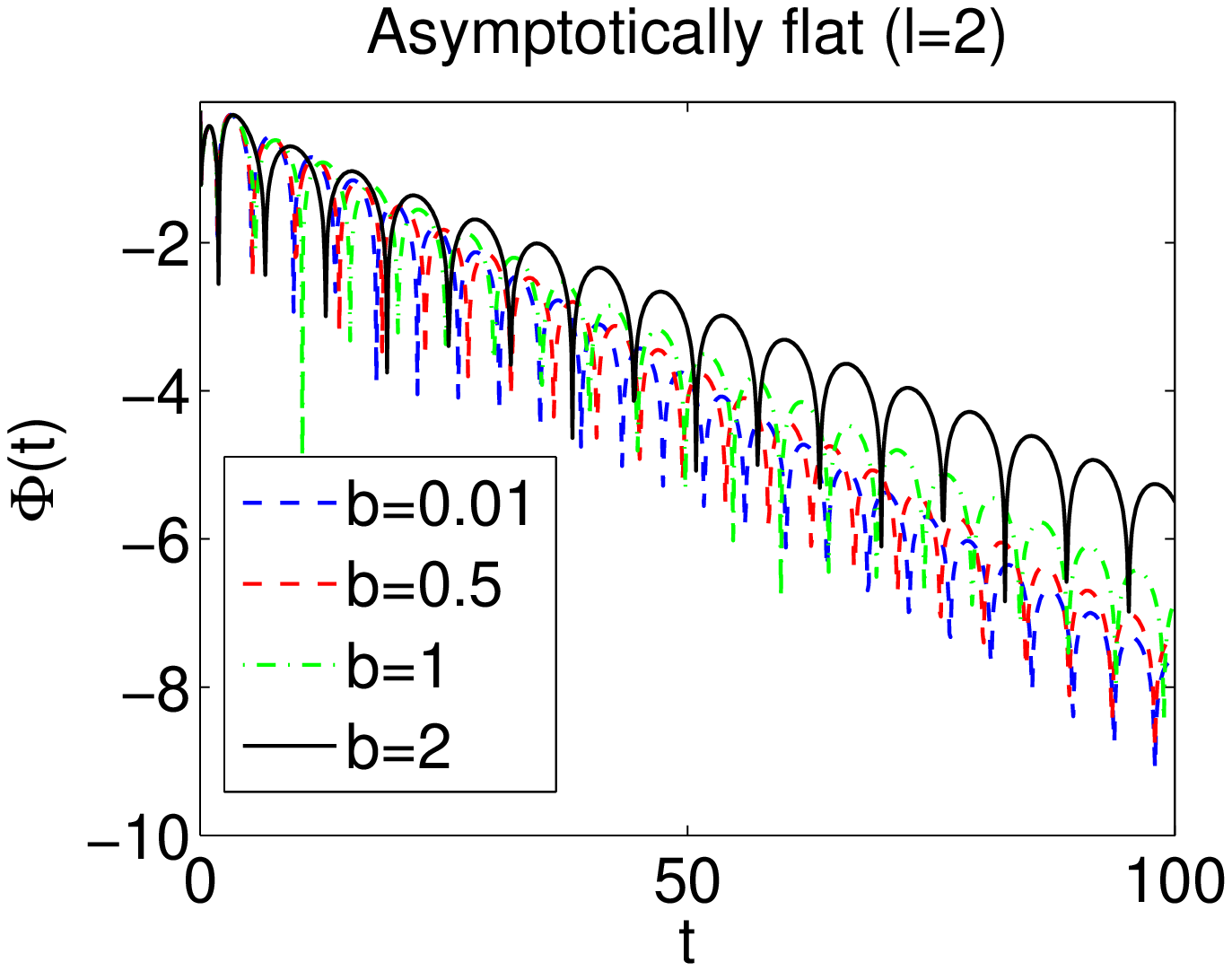}}
\caption{Left panels: the effective potential of even parity perturbation in asymptotically flat spacetime. Right panels: the corresponding temporal evolution of even parity gravitational perturbation field in asymptotically flat spacetime. 
Calculations are carried out by using $r_{\text{h}}=1$.}\label{evenflat}
\end{figure}

\begin{figure}[htbp]
\centerline{\includegraphics[height=4cm]{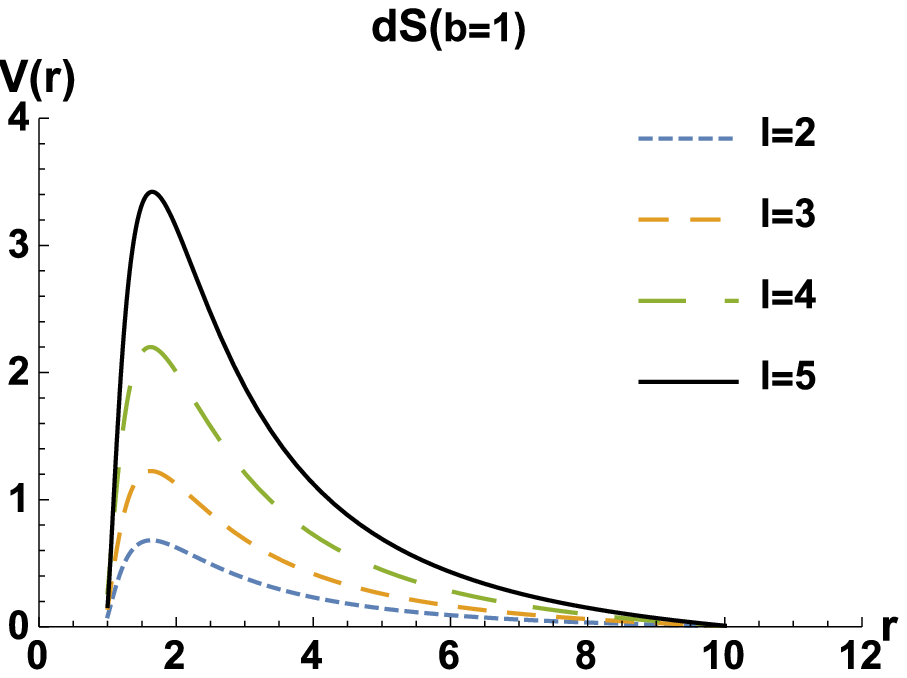}\includegraphics[height=4cm]{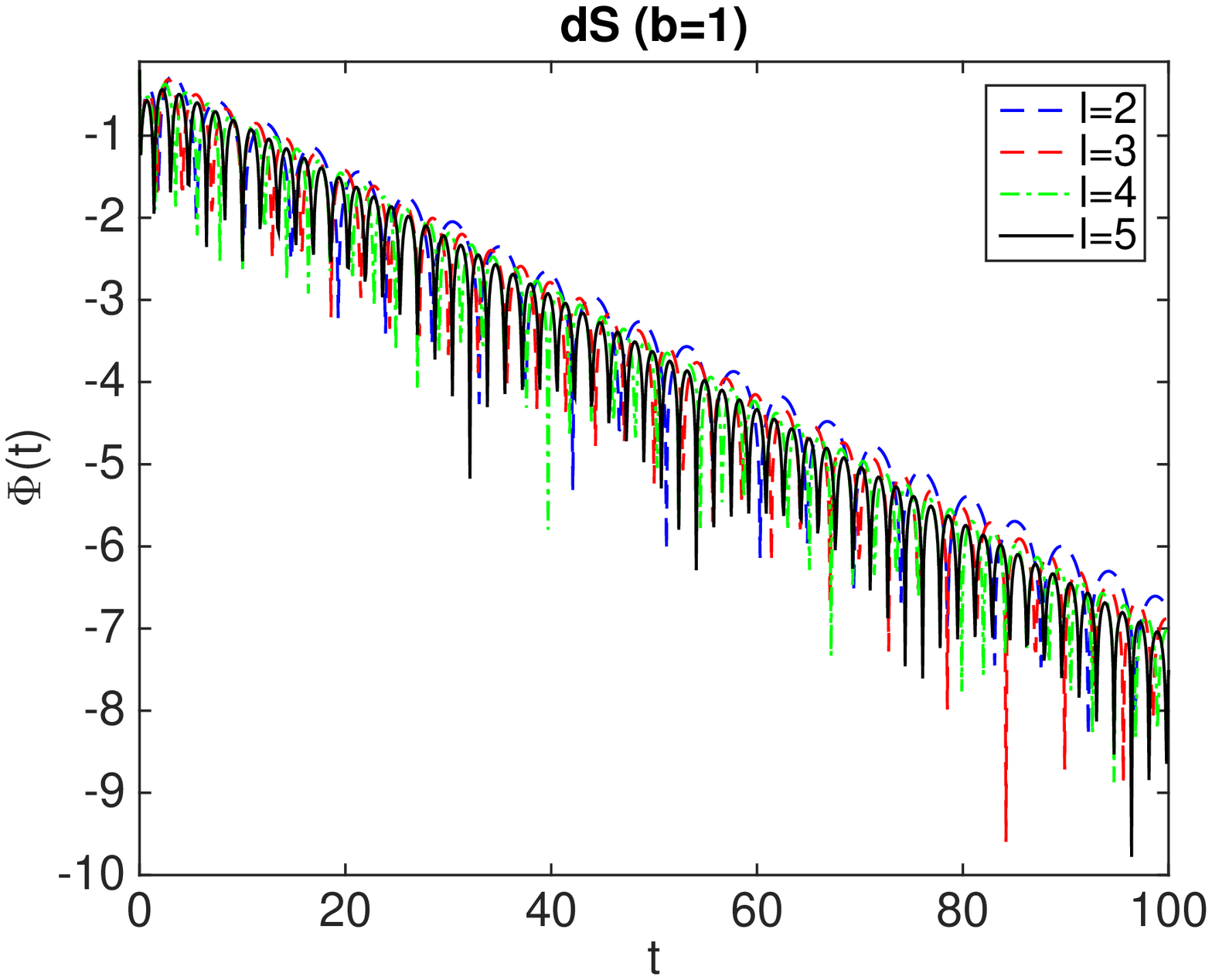}}
\centerline{\includegraphics[height=4cm]{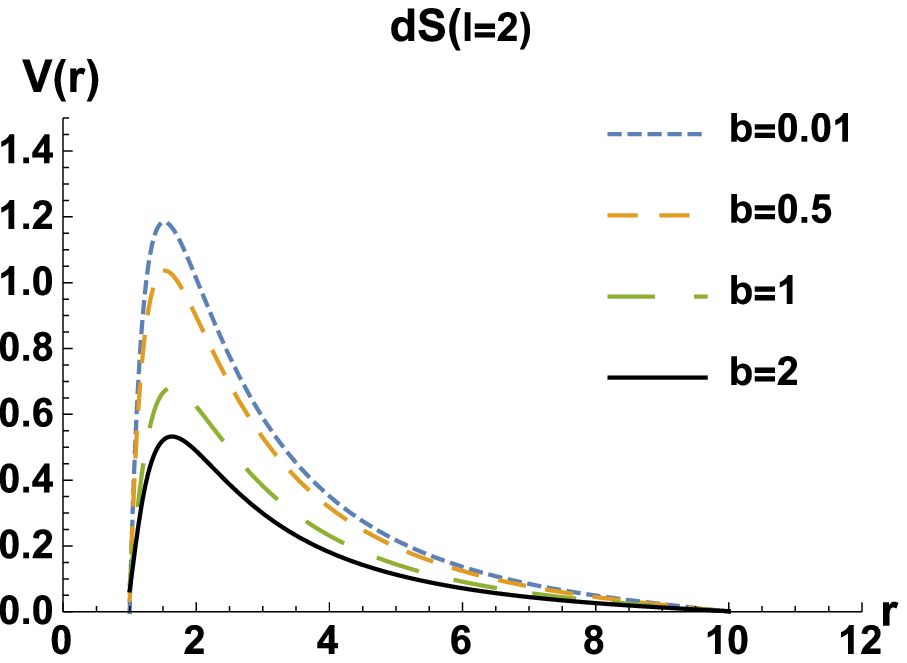}\includegraphics[height=4cm]{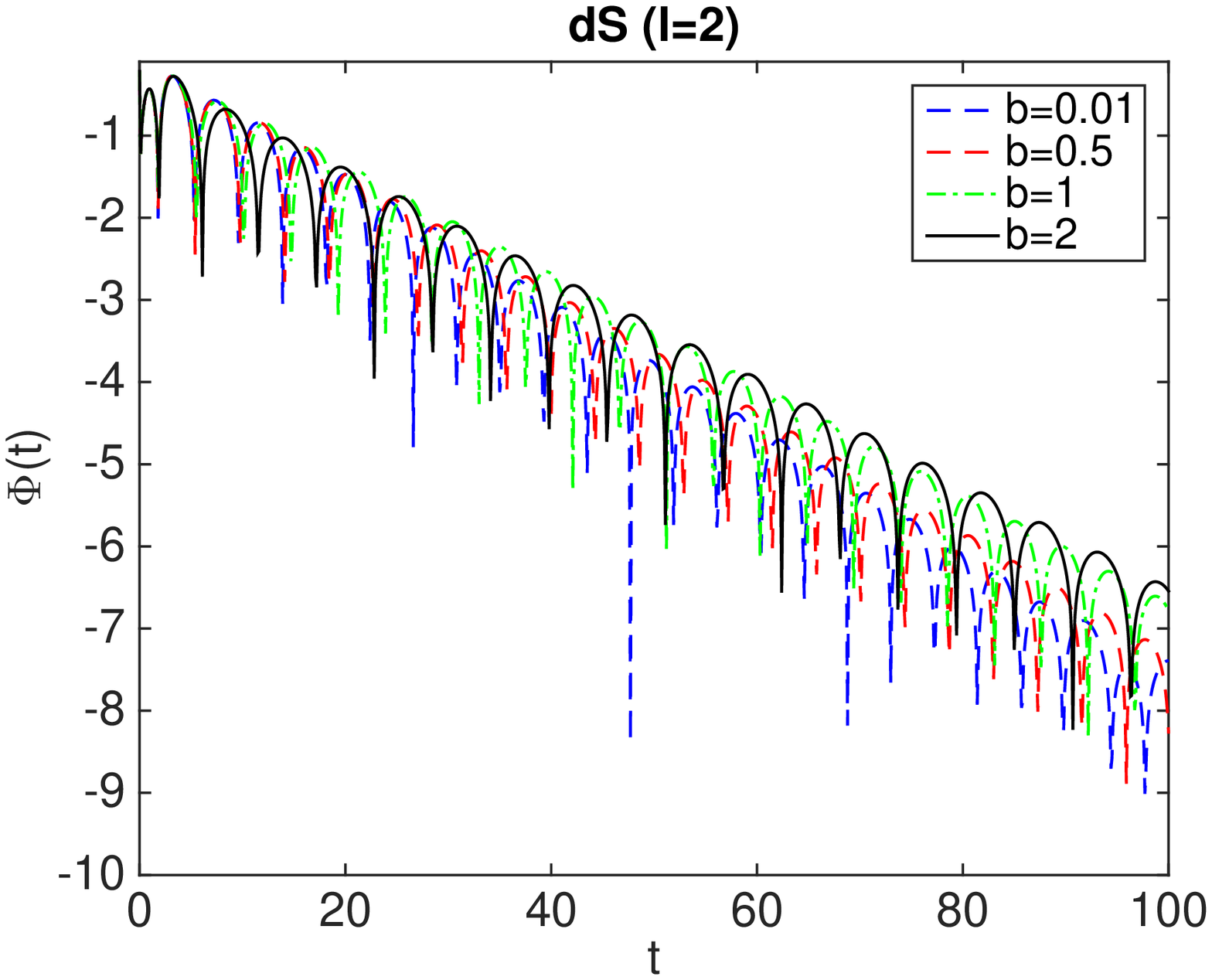}}
\caption{Left panels: the effective potential of even parity perturbation in dS spacetime. Right panels: the corresponding temporal evolution of even parity gravitational perturbation. In the calculations, we choose $r_{\text{h}}=1, r_{c}=10$.}\label{evendS}
\end{figure}

\begin{figure}[htbp]
\centerline{\includegraphics[height=4cm]{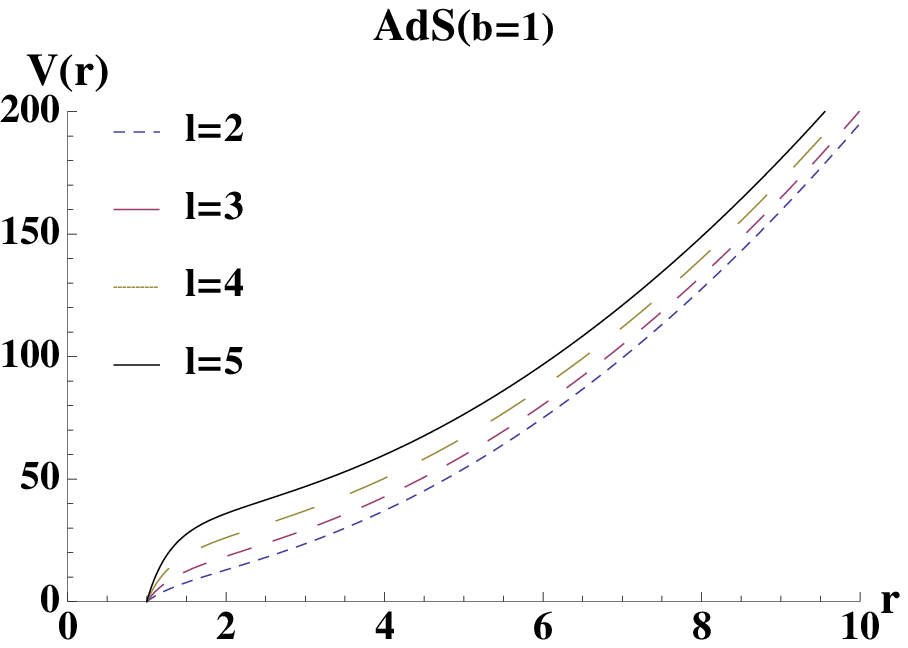}\includegraphics[height=4cm]{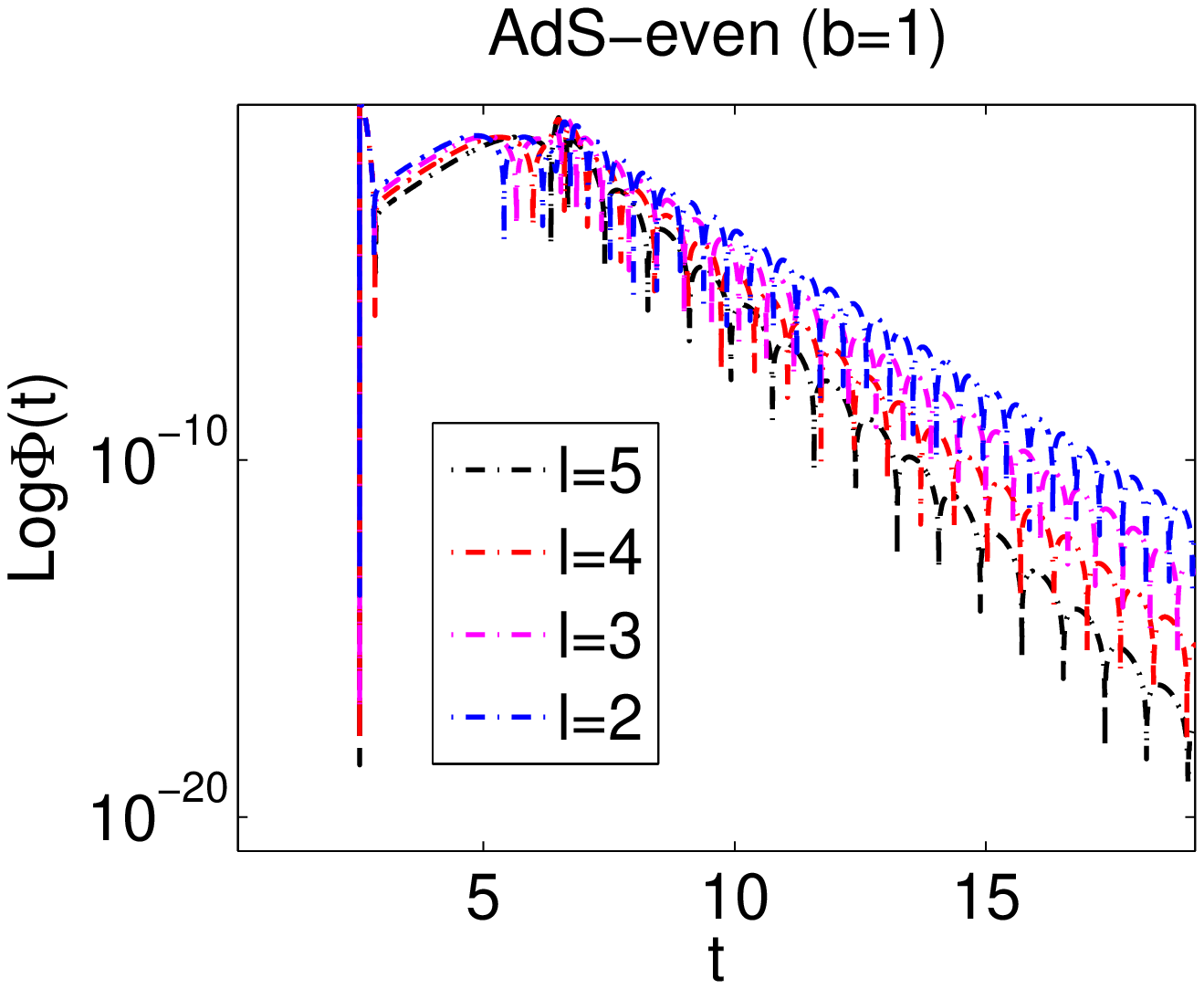}}
\centerline{\includegraphics[height=4cm]{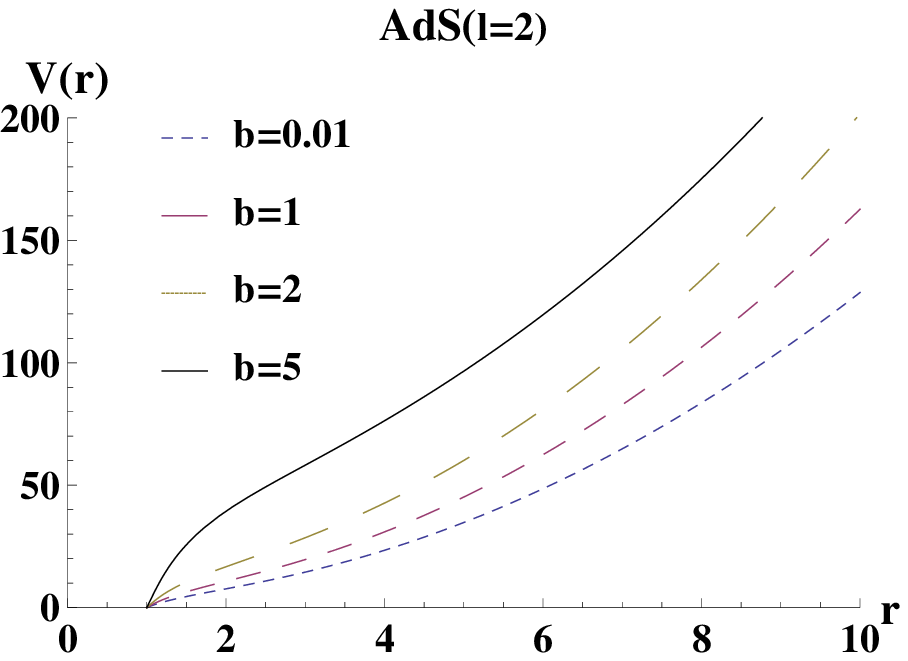}\includegraphics[height=4cm]{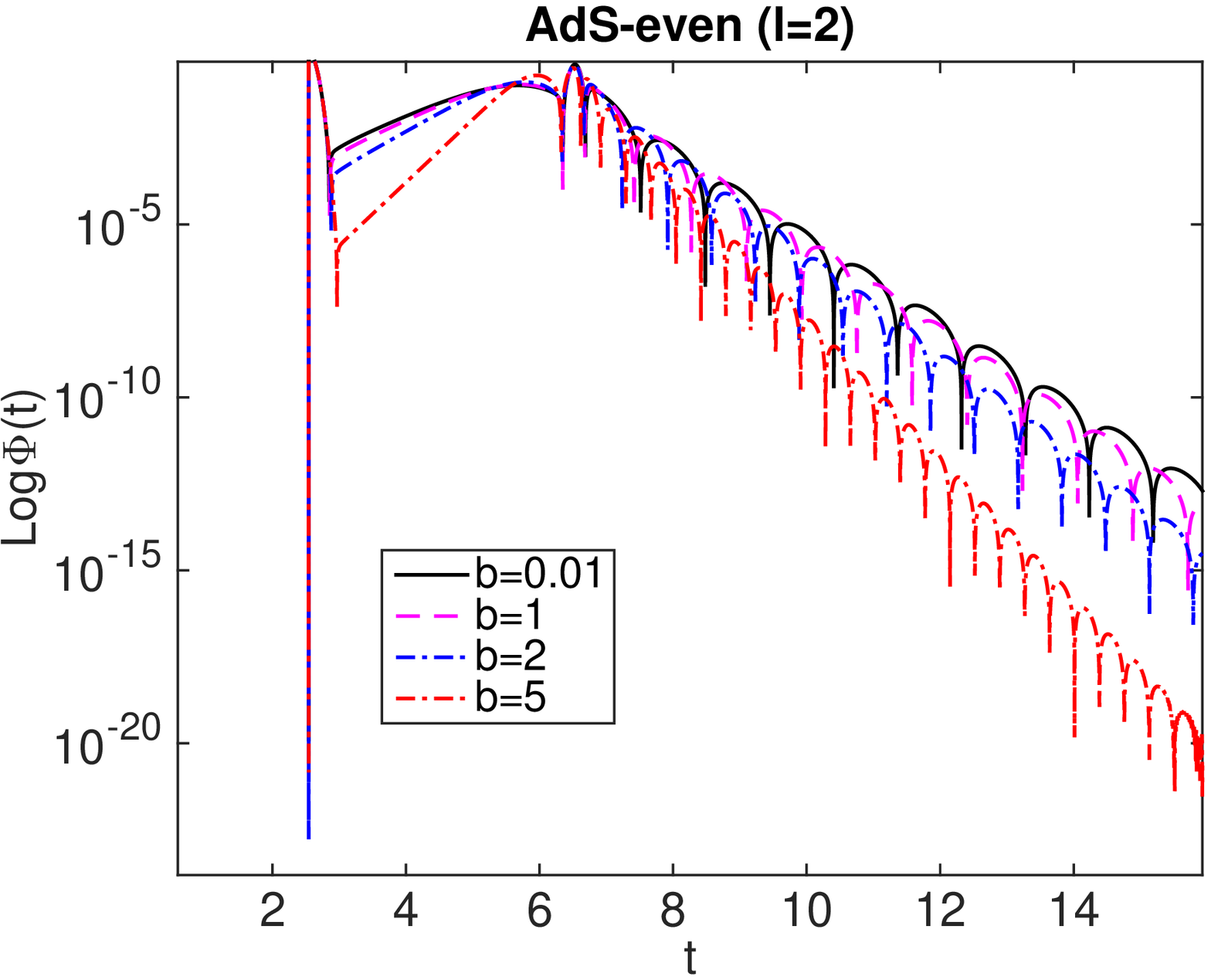}}
\caption{Left panels: the effective potential of even parity perturbation in AdS spacetime. Right panels: the corresponding temporal evolution of even parity gravitational perturbation in AdS spacetime.  In the calculations, we choose  $r_{\text{h}}=1$.}\label{evenadS}
\end{figure}

The resulting master equation can be derived by evaluating
\ba &&
\delta \widehat{G}_{23}+\delta \widehat{G}_{22}=\frac{i\omega H_{1}}{f}+\frac{(-2+l+l^{2})f-2\omega^{2}p^{2}}{2f^{2}pp'}K-\frac{pf'K'}{2fp'}-\frac{-2+l+l^{2}+2p^{2}V(\phi)+2pf'p'}{2fpp'}H_{0}=0.
\ea
By substituting Eq.(\ref{H0a}), (\ref{H1a}) and (\ref{KHa}) into the above equation, the corresponding master equation is given by
\ba &&
\Psi''(r)+\Psi'(r)\left\{\frac{A_{2}}{A_{1}}+\frac{A'_{1}}{A_{1}}+\frac{-2r\chi_{1}-f'}{2f}+\frac{p'(p(2\omega^{2}+2fV(\phi)+r\chi_{1}f'+f'^{2})+2(-r\omega^{2}+ff')p')}{fP_{v}}\right\}
 \nonumber\\ &&
+\Psi(r)\left\{\frac{-r(1+A_{2})\chi_{1}(P_{v}-pf'p')}{A_{1}fP_{v}}+\frac{1}{2A_{1}fP_{v}}(-4r\omega^{2}p'^{2}+A_{2}(-P_{v}f'+p'(p(4\omega^{2}+f'^{2})-4r\omega^{2}p'))
\right.\nonumber\\ && \left.
+2f(P_{v}A'_{2}+2A_{2}p'(pV(\phi)+f'p')))\right\}=0,
\ea
where $P_{v}(r)=-2+l+l^{2}+2p^{2}V(\phi)+3pf'p'$. 
Finally, one defines $\Psi(r)=B_{1}(r)B_{2}(r)\Phi(r)$ and $f(r)=Q(r)F(r)$ (where $Q(r)$ is the coefficient of $\omega^{2}$), the master equation can be simplified into the following equation:
\begin{equation}
\frac{d^{2}\Phi(r)}{dr^{2}_{*}}+\left\{\omega^{2}-V_{e}(r)\right\}\Phi(r)=0,
\end{equation}
where $dr_{*}=dr/F(r)$,
\begin{equation}
V_{e}=\frac{F(r) \left(2 Q(r) \left(F'(r) Q'(r)+F(r) Q''(r)\right)-F(r) Q'(r)^2\right)+4 \tilde{V}(r)}{4 Q(r)^2},
\label{Ve}
\end{equation}
where
\ba &&
Q(r)=\frac{1}{2\sqrt{A_{1} \left(3 p f' p'+l^2+l+2 p^2 V(\phi)-2\right)^2}}\left\{4 f p' \left(r \left(A_{1}'-2\right) p'-p A_{1}'
\right.\right.\nonumber\\ &&\left.\left.
+A_{2} \left(p-r p'\right)\right) \left(3 p f' p'+l^2+l+2 p^2 V(\phi)-2\right)+A_{1} \left(r^2 \chi_{1}^2 \left(2 p f' p'+l^2+l+2 p^2 V(\phi)-2\right)^2
\right.\right.\nonumber\\ &&\left.\left.
+2 \left(r p' \left(4 \left(l^2+l-2\right) f p''-f' p' \left(2 f p'^2+3 \left(l^2+l-2\right)\right)\right)+p^3 \left(V(\phi) \left(6 f' p'-4 f p''\right)
\right.\right.\right.\right.\nonumber\\ &&\left.\left.\left.\left.
+4 f p' \phi ' V'(\phi)\right)+p \left(-2 f \left(3 r f'' p'^3+\left(l^2+l-2\right) p''+2 r p'^3 V(\phi)\right)+f' p' \left(2 f p' \left(3 r p''+p'\right)+3 \left(l^2+l-2\right)\right)
\right.\right.\right.\right.\nonumber\\ &&\left.\left.\left.\left.
-8 r f'^2 p'^3\right)+2 p^2 p' \left(f \left(p' \left(3 f''-2 r \phi ' V'(\phi)\right)+2 V(\phi) \left(2 r p''+p'\right)\right)+f' p' \left(4 f'-3 r V(\phi)\right)\right)\right)\right)
\right\}^{1/2},
\ea

\ba &&
\tilde{V}(r)=\frac{1}{16 A_{1}^2 \left(l^2+l+2 p^2 V(\phi)+3 p f' p'-2\right)^2}
\left(-16 \left(p^2 \left(p'^2-2 p p''\right) V(\phi)^2+\left(2 p f' p'^3-\left(l^2+l-p^2 f''-2\right) p'^2
\right.\right.\right.\nonumber\\ &&\left.\left.\left.
-4 p^2 f' p'' p'-\left(l^2+l-2\right) p p''\right) V(\phi)+p' \left(\left(2 p'^3-3 p p' p''\right) f'^2-\left(p' V'(\phi) \phi ' p^2+2 \left(l^2+l-2\right) p''\right) f'
\right.\right.\right.\nonumber\\ &&\left.\left.\left.
-\left(l^2+l-2\right) \left(p V'(\phi) \phi '+p' f''\right)\right)\right) f^2-4 \left(12 V(\phi)^2 f'' p^4+2 f' \left(2 p' V(\phi)^2+\left(16 p' f''-f' p''\right) V(\phi)
\right.\right.\right.\nonumber\\ &&\left.\left.\left.
+f' p' V'(\phi) \Psi '\right) p^3+2 \left(12 f'^2 f'' p'^2+V(\phi) \left(5 f'^2 p'^2+6 \left(l^2+l-2\right) f''\right)\right) p^2+f' \left(4 f'^2 p'^3+2 \left(l^2+l-2\right) V(\phi) p'
\right.\right.\right.\nonumber\\ &&\left.\left.\left.
+16 \left(l^2+l-2\right) f'' p'-\left(l^2+l-2\right) f' p''\right) p+\left(l^2+l-2\right) f'^2 p'^2+3 \left(l^2+l-2\right)^2 f''\right) f
\right.\nonumber\\ &&\left.
+\left(8 p p' f'^2+3 \left(l^2+l+2 p^2 V(\phi)-2\right) f'\right)^2\right) A_{1}^2+4 A_{2} f \left(l^2+l+2 p^2 V(\phi)+3 p f' p'-2\right) \left(2 p p' f'^2
\right.\nonumber\\ &&\left.
+\left(l^2+l-4 f p'^2+2 p^2 V(\phi)-2\right) f'-4 f p V(\phi) p'\right) A_{1}-4 f \left(l^2+l+2 p^2 V(\phi)+3 p f' p'-2\right)
\nonumber\\ &&
\left(A_{1}' f' \left(l^2+l+2 p^2 V(\phi)+2 p f' p'-2\right)+2 f \left(-A_{1}'' l^2-A_{1}'' l-2 A_{1}' f' p'^2+A_{2}'(r) \left(l^2+l+2 p^2 V(\phi)
\right.\right.\right.\nonumber\\ &&\left.\left.\left.
+3 p f' p'-2\right)-2 p^2 V(\phi) A_{1}''+2 A_{1}''-p p' \left(2 V(\phi) A_{1}'+3 f' A_{1}''\right)\right)\right) A_{1}
\nonumber\\ &&
+4 A_{2}^2 f^2 \left(l^2+l+2 p^2 V(\phi)+3 p f' p'-2\right)^2-4 f^2 A_{1}'^2 \left(l^2+l+2 p^2 V(\phi)+3 p f' p'-2\right)^2.
\ea
Here we study the QNMs for even parity perturbation in asymptotically flat, dS and AdS spacetimes. 
Fig.\ref{evenflat}, \ref{evendS} and \ref{evenadS} show the effective potential functions and corresponding temporal evolution of even parity perturbation in asymptotically flat, dS and AdS spacetimes respectively. 
We consider the potential in the range $r>r_{h}$ in asymptotically flat and AdS spacetimes, and $r_{h}<r<r_{c}$ in dS spacetime. 
Here, the potential functions are all positive definite so that the corresponding regular phantom BHs are likely to be stable. 
Furthermore, in asymptotically flat and dS spacetimes, the relationships between $\text{Re}(\omega)$, $\text{Im}(\omega)$ and $b$, $l$ are similar to those of odd parity, the differences are in details. 
However, the potential function in AdS spacetime is mostly a convex function, and therefore it approaches infinity faster than that of odd parity. 
Nevertheless, the resulting QNMs in AdS spacetime are also found to be stable.

\section{Hawking radiation of the regular phantom black hole}
\label{4}
As mentioned before in section {\ref{2}}, the interior structures of regular BHs are quite different from those of traditional ones. 
In order to achieve a better understanding of the properties of such regular phantom BHs, it is meaningful to investigate the corresponding Hawking radiation, which is considered to be a promising method to study the thermodynamics of BHs.

From a quantum mechanics viewpoint, QNM carries information on the quantization of the system around the BH's horizon \cite{Hawking1,Hawking2,Hawking3,Hawking5,Hawking7}. 
The Hawking radiation spectrum provides an interpretation of such quantization at an effective temperature \cite{Hawking1,Hawking2,Hawking8}. 
In this paper, we adopt a simple and effective method (i.e.,Hamilton-Jacobi equation) to analyze the Hawking radiation of the regular phantom BH. 
Based on the tunneling theory of Hawking radiation \cite{addHJ2,addHJ3}, this method was first put forward by Srinivasan in 1999 \cite{addHJ1}, and plays an important role in recent studies \cite{addHJ4,addHJ5}.

Up to this point, we have been using the natural units in the calculations of QNM. 
However, since the Planck constant $\hbar$ is very important in the study of tunneling radiation, we will explicitly write it down here in the field functions. 
In other words, $h_{0}$ and $h_{1}$ of Eq.(\ref{eqh}) are now rewritten as $h_{0}(r,t)=\exp(-i\omega t/\hbar)h_{0}(r)$, $h_{1}(r,t)=\exp(-i\omega t/\hbar)h_{1}(r)$, and Eq.(\ref{evenper}) becomes
\begin{equation}
h_{\mu\nu}=\exp{(-i\omega t/\hbar)}\left(
\begin{array}{cccc}
 H_{0}(r)f(r) & H_{1}(r) & 0 & 0 \\
H_{1}(r) & H_{2}(r)/f(r) & 0 & 0\\
 0 & 0 & p(r)^{2}K(r) & 0\\
 0 & 0 & 0 & p(r)^{2}K(r)\sin^{2}\theta
\end{array}
\right)P_{l}(cos\theta).
\label{evenper2}
\end{equation}
The corresponding Schr$\ddot{o}$dinger-type equation can be expressed as
\begin{equation}
\frac{d^{2}\Phi}{dr_{*}^{2}}+(\frac{\omega^{2}}{\hbar^{2}}-V(r))\Phi=0.
\label{masterHK}
\end{equation}

Since the Hawking radiation reflects BH's radial properties in the vicinity of the horizon, it is worthwhile to explicitly discuss radial field equations. 
According to Eq.(\ref{masterHK}), near the horizons $r\rightarrow r_{\text{h}}$, one has $V_{o}\rightarrow0$ and $V_{e}\rightarrow0$, so that the field equation can be simplified to
\begin{equation}
\frac{d^{2}\Phi(r)}{dr^{2}_{*}}+\frac{\omega^{2}}{\hbar^{2}}\Phi(r)=0.
\label{Hawkingeq}
\end{equation}
By introducing $\Phi(r)=Ce^{-\frac{i}{\hbar}R(r)}$ and using the semi-classical approximation to neglect the terms of $\mathcal{O}(\hbar)$, as $\hbar$ is small, one obtains
\begin{equation}
-f^{2}R'^{2}+\omega^{2}=0.
\end{equation}
Therefore, $R'(r)=\pm\omega/f$, where $\pm$ represents ingoing or outgoing mode. 
The above equation is no other than the Hamilton-Jacobi equation at the horizon.
We note that the above argument also applies to the case of asymptotically dS spacetime, where the event horizon $r_{\text{h}}$ and cosmological horizon $r_{\text{c}}$ should be both considered.
In fact, according to the Hamilton-Jacobi equation $g^{\mu\nu}\frac{\partial S}{\partial x^{\mu}}\frac{\partial S}{\partial x^{\nu}}+m^2=0$, the radial Hamilton-Jacobi equation can be transformed into Eq.(49) at the horizon, so we can use this result to study the Hawking tunneling radiation under semi-classical approximation \cite{KL1,KL2,KL3}.

Owing to the coordinate singularity at the horizon, the integration from inner surface of horizon to outer surface shall be carried out by using Residue Theorem:
\begin{equation}
R(r_{\text{h}})=\pm\frac{i\pi\omega}{f'(r_{\text{h}})}.
\end{equation}
Then, the tunneling rate of Hawking radiation is \cite{Hawking9,Hawking10}
\begin{equation}
\Gamma=\frac{\exp{(-2\text{Im}(R_{+})})}{\exp{(-2\text{Im}(R_{-})})}=\exp{(-\frac{4\pi\omega}{f'(r_{\text{h}})})}.
\end{equation}
According to the relationship between tunneling rate $\Gamma$ and the Hawking temperature, the temperature of the BH in the vicinity of the event horizon $r_{\text{h}}$ is
\begin{equation}
T_{H}=\frac{f'(r_{\text{h}})}{4\pi}.
\end{equation}
From the inside of the cosmological horizon $r_{\text{c}}$ in dS spacetime, we consider the incident wave solution of Hawking radiation. 
Therefore, there is an extra minus sign in the resulting expression of the Hawking temperature in the neighborhood of the cosmological horizon \cite{Hawking11} as follows:
\begin{equation}
T_{c}=-\frac{f'(r_{\text{c}})}{4\pi}.
\end{equation}
\begin{figure}[htbp]
\centerline{\includegraphics[height=4cm]{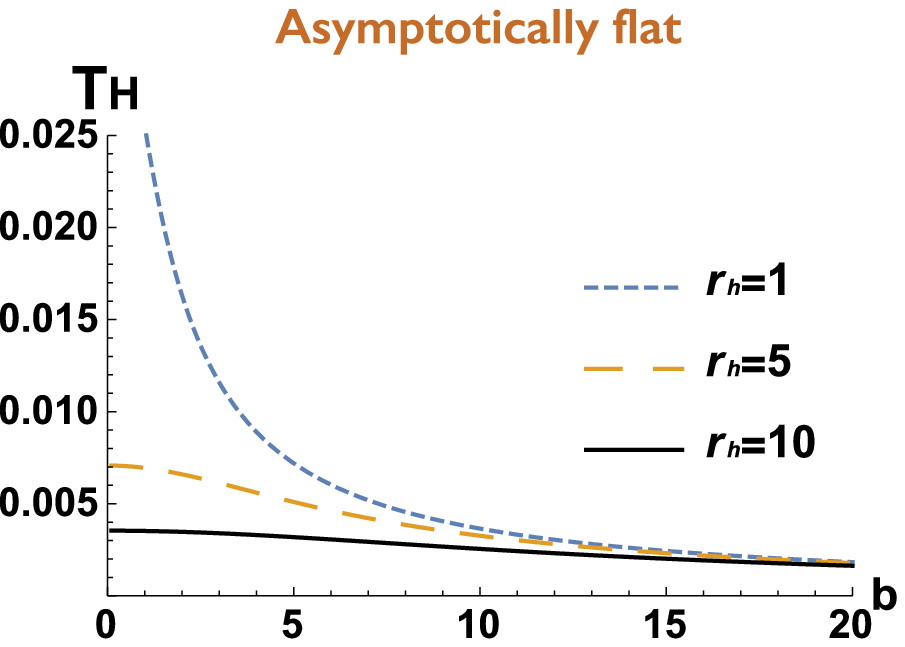}\includegraphics[height=4cm]{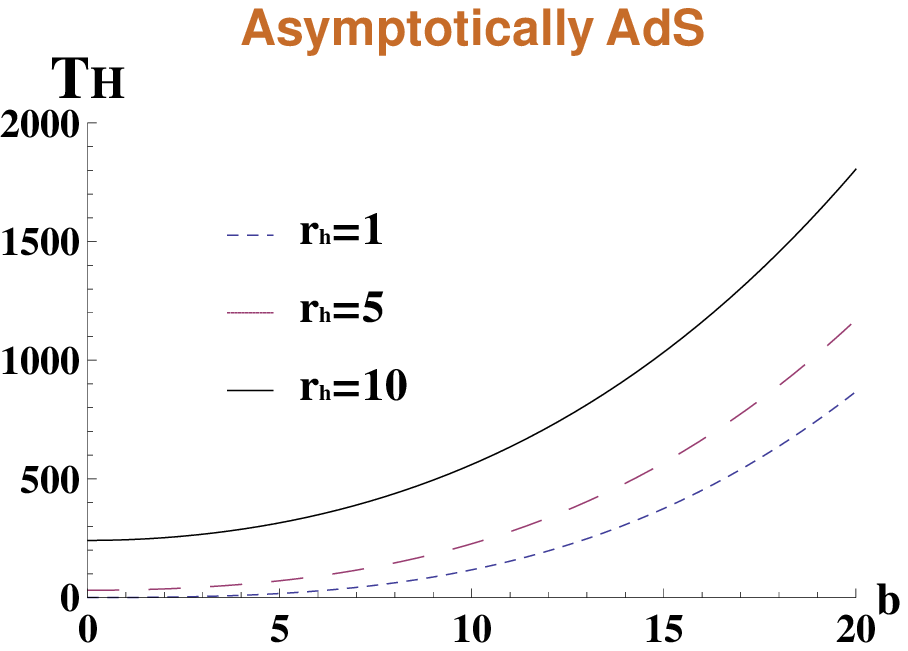}}
\centerline{\includegraphics[height=4cm]{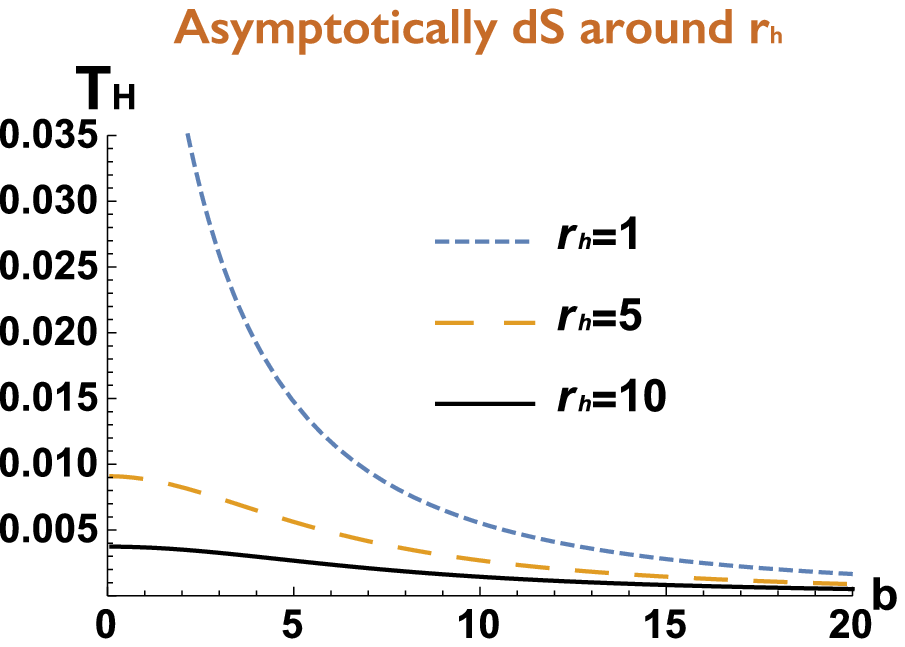}\includegraphics[height=4cm]{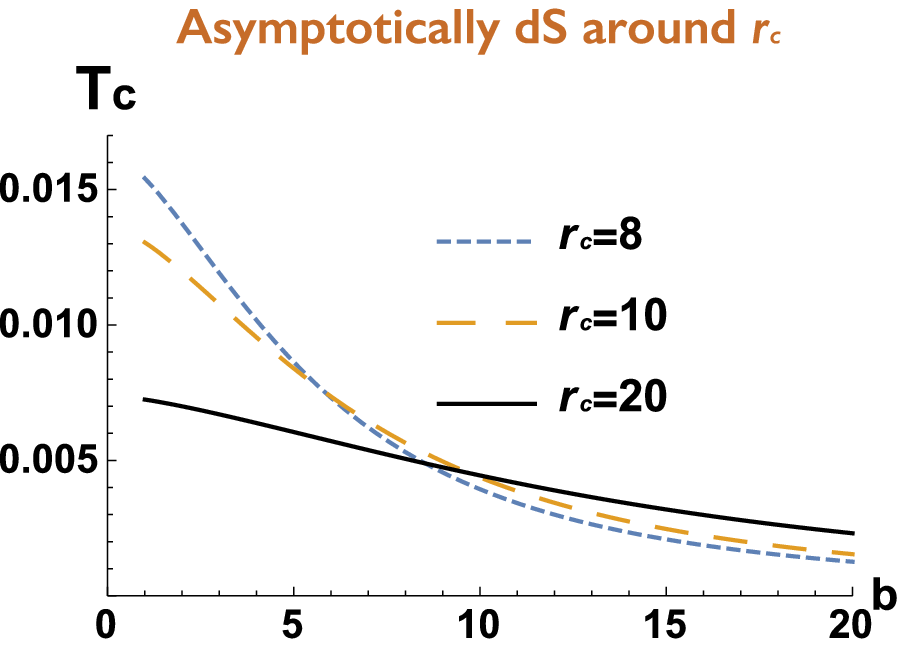}}
\caption{The Hawking temperature as a function of $b$ in asymptotically flat and dS spacetimes.}\label{Hawking}
\end{figure}

Fig.\ref{Hawking} shows the impact of the parameter $b$ on the thermodynamics of a regular phantom BH for different spacetimes. 
In asymptotically flat spacetime, the Hawking temperature $T_{H}$ decreases monotonously with increasing $b$ and a given $r_h$; and for a given $b$, $T_{H}$ becomes smaller with increasing $r_{h}$. 
In asymptotically AdS spacetime, however, the Hawking temperature $T_{H}$ increases monotonously with increasing $b$ and a given $r_h$; and for a given $b$, $T_{H}$ becomes larger with increasing $r_{h}$. 
In asymptotically dS spacetime, the relationship between $T_{H}$ and the event horizon $r_{h}$ is similar to that in asymptotically flat case;
while the Hawking temperature $T_{C}$ in the vicinity of the cosmological horizon $h_{c}$ also decreases monotonously with increasing $b$, and it decays faster than other cases. 
The different dependence of Hawking temperature on $b$ probably results from the distinctive spacetime structure of AdS phantom BH from others.

\section{Conclusions and Remarks}
\label{5}

We studied static spherically symmetric solutions of regular phantom BH in asymptotically flat, dS, and AdS spacetimes, and then investigated the QNMs of gravitational perturbations as well as Hawking radiations for these BH spacetimes. 
In our calculations, besides the metric perturbation $\delta g_{\mu\nu}$ of the Ricci curvature tensor and scalar curvature, its effect on the energy momentum tensor of the matter field is also considered. 
In the derivation of the master equation for even parity, we made use of the method proposed in \cite{HD1}. 
However, the BH metric in this paper is a self-gravitating solution of a minimally coupled scalar field with an arbitrary potential rather than the Lovelock equations, so that it does not satisfy the relation of Eq.(5.4) in \cite{HD1}. 
As a result, the obtained master equation is more complicated.
It is found that the calculated effective potential of AdS spacetime in the limit $r\rightarrow\infty$ is very different from those of asymptotically flat as well as dS spacetimes. 
For the asymptotical AdS spacetime, the effective potentials diverge at infinity, which implies that the wave function $\Phi$ should vanish in this limit \cite{AdS1,AdS2,AdS3,AdS4,AdS5}. 
At the outside of the event horizon, on the other hand, the corresponding wave function must be an incoming wave. 
The distinct nature of AdS BH spacetime leads to that some traditional numerical methods such as WKB approximation, continuous fraction method cease to be valid. 
In our calculations, we therefore employed the finite difference method to numerically calculate the temporal evolution of small gravitational perturbations.
The relationships between Hawking temperature and parameters such as $b, r_{\text{h}}$ in different spacetime are also studied.

Owing to the importance of the parameter $b$, which carries the physical content of the coupling strength between phantom scalar field and the gravity, 
we studied the dependence of various physical quantities on this specific parameter, among others.
From the above calculated results, we draw the following conclusions.
For odd parity, in asymptotic flat and dS spacetime, the perturbation frequency ($\text{Re}(\omega)$) becomes smaller and the decay rate ($|\text{Im}(\omega)|$) decreases when $b$ increases.
The perturbation frequency ($\text{Re}(\omega)$) and the decay rate ($|\text{Im}(\omega)|$) both increase with increasing $l$. 
For even parity, the shape of the effective potential in the vicinity of the event horizon is very different from that for odd parity.
In particular, $V_{e}(r)$ decreases monotonically in the range $r>r_{h}$, which is caused by the metric transformation $F(r)=f(r)/\text{Q}(r)$. 
However, our numerical calculations show that the resultant gravitational perturbations are stable in all three different spacetime backgrounds.
The Hawking temperature $T_{H}$ (and $T_{\text{c}}$) of the phantom BH decrease with increasing $b$ in asymptotically flat and dS spacetimes.
On the contrary, in AdS spacetime, the temperature $T_{H}$ increases with increasing $b$. 
These results reflect distinct natures of the horizons of different spacetimes. 

Recently, another type of regular phantom BH, which is a solution of Einstein-Maxwell equations, was proposed by Lemos and Zanchin\cite{Lemos}. 
The properties of interior as well as exterior structures were investigated. 
It is meaningful to investigate the stability of such BH solution and compare it with the phantom BH in the present paper.

\section*{Appendix}
We note that, $V(\phi(r))$ in Eq.(\ref{Vr}) is just the 0-order solution. 
However, when we investigate the perturbation in the BH background, the potential could, in principle, be expanded to higher order terms.
Since the higher order terms of $V(\phi)$ have not been determined yet, we can freely choose the form of such higher order terms to cancel out the contributions of the perturbation of phantom scalar field.
In the Appendix, we carry out explicit calculations to show this is indeed possible.

We consider the following metric and phantom scalar field
\begin{equation}
g_{\mu\nu}=\bar{g}_{\mu\nu}+h_{\mu\nu},~~~~~\phi=\bar{\phi}+\phi_{1},
\end{equation}
where $\bar{g}_{\mu\nu}$, $\bar{\phi}$ are background metric and phantom scalar field respectively, and $\phi_{1}$ represents the first order perturbation of phantom scalar field. Then
\begin{equation}
g^{\mu\nu}=\bar{g}^{\mu\nu}-h^{\mu\nu}+h^{\mu}_{\alpha}h^{\nu\alpha}.
\end{equation}
The corresponding metric and potential of phantom scalar field can be expressed as
\begin{equation}
\sqrt{-g}=\sqrt{-\bar{g}}+\sqrt{-g_{1}}+\sqrt{-g_{2}}+\cdot\cdot\cdot,
\end{equation}
\begin{equation}
R=\bar{R}+R_{1}+R_{2}+\cdot\cdot\cdot~~~~~V(\phi)=\bar{V}+V_{1}+V_{2}+\cdot\cdot\cdot,
\end{equation}
where the superscript $^{-}$ and subscrips $_{1}$,$_{2}$ represent the background quantities, the first, and the second order perturbations respectively.
Therefore, the action of the gravity with the phantom scalar field, Eq.(\ref{action}), can be expanded as
\begin{equation}
S=\bar{S}+S_{1}+S_{2}+\cdot\cdot\cdot,
\end{equation}
where
\begin{equation}
\bar{S}=\int d^{4}x\sqrt{-\bar{g}}(\bar{R}+\epsilon \bar{g}^{\mu\nu}\bar{\phi}_{,\mu}\bar{\phi}_{,\nu}-2\bar{V}),
\end{equation}
\begin{equation}
S_{1}=\int d^{4}x\left\{\sqrt{-\bar{g}}[R_{1}+\epsilon(2\bar{\phi}_{,\mu}\phi^{,\mu}_{1}-h^{\mu\nu}\bar{\phi}_{,\mu}\bar{\phi}_{,\nu})-2V_{1}]+\sqrt{-g_{1}}(\bar{R}+\epsilon \bar{g}^{\mu\nu}\bar{\phi}_{,\mu}\bar{\phi}_{,\nu}-2\bar{V})\right\},
\end{equation}
\ba &&
S_{2}=\int d^{4}x\left\{\sqrt{-\bar{g}}[R_{2}+\epsilon(\phi_{1,\mu}\phi_{1}^{,\mu}-2h^{\mu\nu}\bar{\phi}_{,\mu}\phi_{1,\nu}+h^{\mu}_{\alpha}h^{\nu\alpha}\bar{\phi}_{,\mu}\bar{\phi}_{,\nu})-2V_{2}]
\right.\nonumber\\ &&\left.
+\sqrt{-g_{1}}[R_{1}+\epsilon( 2\bar{\phi}_{,\mu}\phi_{1}^{,\mu}-h^{\mu\nu}\bar{\phi}_{,\mu}\bar{\phi}_{,\nu})-2V_{1}]+\sqrt{-g_{2}}(\bar{R}+\epsilon\bar{g}^{\mu\nu}\bar{\phi}_{,\mu}\bar{\phi}_{,\nu}-2\bar{V})\right\}.
\ea
In the background equation, there is no effect of $V_{1}$ and $V_{2}$, while in the perturbed equation we can always choose the form of the perturbed potential $V_{1}$, $V_{2}$ as
\begin{equation}
V_{1}=\epsilon\bar{\phi}_{,\mu}\phi^{,\mu}_{1},~~~~~V_{2}=\frac{\epsilon}{2}\phi_{1,\mu}\phi_{1}^{,\mu}-\epsilon h^{\mu\nu}\bar{\phi}_{,\mu}\phi_{1,\nu},
\end{equation}
to cancel out the contributions from the phantom scalar field perturbation $\phi_{1}$. 
Therefore, in the text, we do not explicitly consider the perturbation of the phantom field.
\\
\\
\section*{ Acknowledgements}
This work is supported by Natural Science Foundation of China Grants No. 11205254, No. 11178018, No.11573022, No. 11605015 and No.11375279, and the Fundamental Research Funds for the Central Universities 106112016CDJXY300002 and 106112015CDJRC131216, and Chongqing Postdoctoral Science Foundation (No.Xm2015027), and the Open Project Program of State Key Laboratory of Theoretical Physics, Institute of Theoretical Physics, Chinese Academy of Sciences, China (No. Y5KF181CJ1). We also acknowledge the financial support by Brazillian fundations 
Funda\c{c}\~ao de Amparo \`a Pesquisa do Estado de S\~ao Paulo (FAPESP), 
Funda\c{c}\~ao de Amparo \`a Pesquisa do Estado de Minas Gerais (FAPEMIG), 
Conselho Nacional de Desenvolvimento Cient\'{\i}fico e Tecnol\'ogico (CNPq),
Coordena\c{c}\~ao de Aperfei\c{c}oamento de Pessoal de N\'ivel Superior (CAPES) and Chinese state scholarship.


.
\end{document}